\documentclass{cernyrep}
\usepackage[T1]{fontenc}
\usepackage[bookmarks, colorlinks=true, linktoc=page, pdftex, linkcolor=black, citecolor=black, urlcolor=blue]{hyperref}

\usepackage{fancyhdr}
\fancyhfoffset{4 mm}
\fancypagestyle{ARTTITLE}{%
\fancyhf{} 
\lhead{Proceedings of the CERN--Accelerator--School course on
{\it Introduction to Accelerator Physics}}
\lfoot{Available online at \url{https://cas.web.cern.ch/previous-schools}}
\rfoot{\thepage\hspace*{3mm}}
 
}

\usepackage{varwidth}
\usepackage{xcolor}

\begin{document}
\title{Warm Magnets}
\author{Gijs de Rijk}
\institute{CERN, Geneva, Switzerland}

\begin{abstract}
Warm magnets are magnets that function in normal ambient temperature conditions. These types are mostly using a soft steel yoke for field amplification and either Copper or Aluminium coils or permanent magnets to generate the field. Magnets powered with such normal-conducting coils are often also called classical, iron dominated or resistive magnets. Since decades these types of magnets are the workhorse for most linear and circular accelerators and beam transfer lines.
\end{abstract}

\maketitle
\thispagestyle{ARTTITLE}

\section{Introduction: magnetic fields and basic magnets}
\subsection{Magnetostatics}
A magnet at constant field can be described by the Maxwell equations with the time dependent terms set to zero; the case of magnetostatics. Let's have a closer look at the three equations that describe magnetostatics (in the differential form): \newline 
Gauss law of magnetism: \hspace{5.9cm} \begin{math} div \overrightarrow{B} = 0 \end{math} \hfill(always holds); \newline 
Ampère's law with no time dependencies:  \hspace{3.35cm}  \begin{math} rot \overrightarrow{H} = \overrightarrow{J} \end{math}  \hfill(magnetostatics); \newline 
Relation between the magnetic field $\overrightarrow{H}$ and the flux density  $\overrightarrow{B}$:   \hspace{0.1cm}   \begin{math} \overrightarrow{B} =  \mu_{0} \mu_{r} \overrightarrow{H} \end{math}       \hfill(linear materials).

\subsection{Iron-dominated magnets}

\begin{figure}[h!]
\begin{center}
\includegraphics[height=2.5cm]{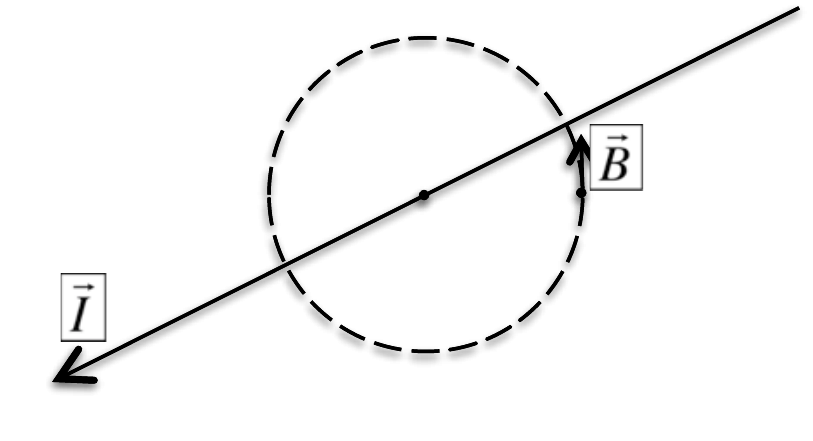} \hspace{ 2cm}
\includegraphics[height=2.5cm]{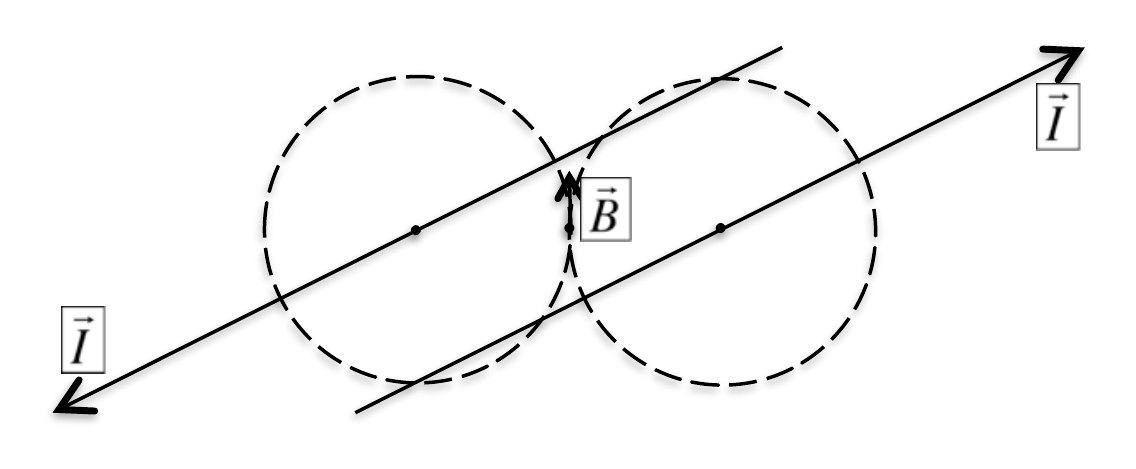}
\caption{left: infinite wire with circular field line; right: two infinite wires with opposite currents.}
\label{fig:current-wires}
\end{center}
\end{figure}

From Ampère's law with no time dependencies (integral form), $ \oint_c \overrightarrow{H}\cdot \overrightarrow{{\rm d}l}  =  I $ we can derive the field generated by a current carrying line (see Fig.~\ref{fig:current-wires} left):  $ \overrightarrow{B} = \frac{\mu_{0}I}{2 \pi r} \cdot \hat{\varphi} $  with $\hat{\varphi} $ the direction vector as the tangent of a circle.
 
If we want to make a  $B=1.5 \UT$ magnet with just two infinitely long thin wires placed at a distance of $100 \Umm$ in air (see Fig.~\ref{fig:current-wires} right) we need $I = 187500 \UA$. We see that to reach reasonable fields ($ \geq 1 \UT$) we need large currents, moreover in such configurations the field quality will be poor.

\begin{figure}[ht!]
\begin{center}
\includegraphics[width=5.8cm]{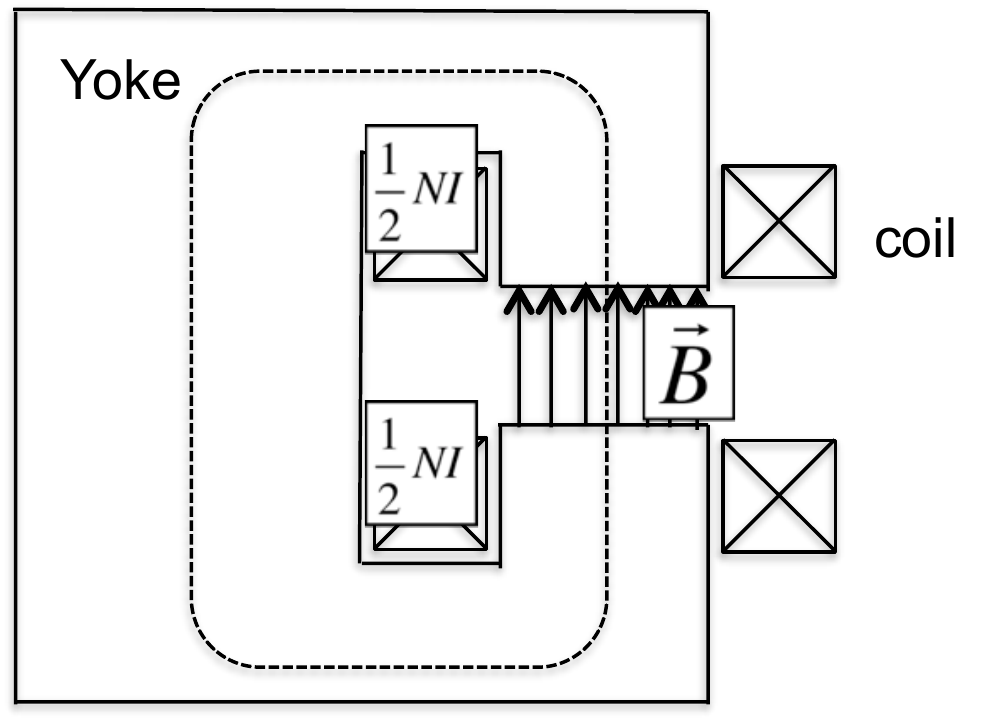} \hspace{ 1.2cm}
\includegraphics[width=7cm]{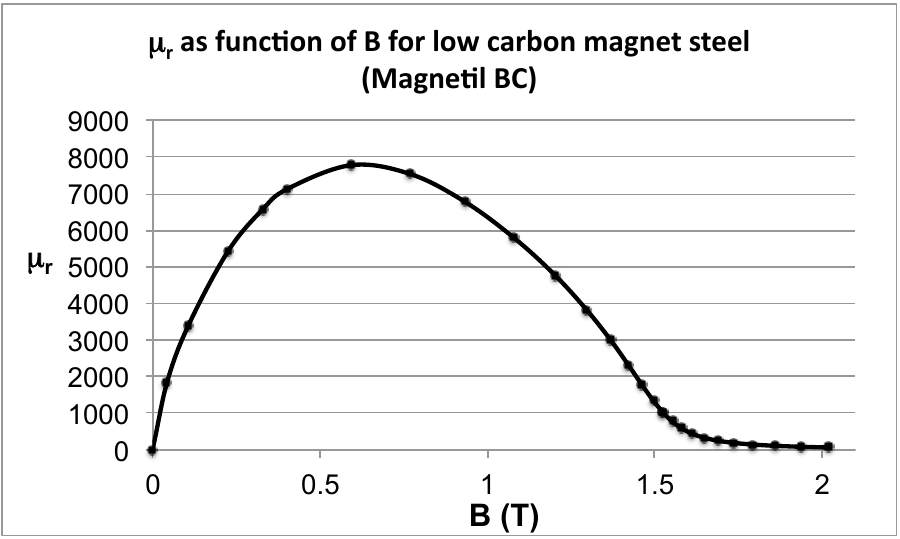}
\caption{left: Iron-dominated magnet: the dotted line is the path-integral loop; each coil carries a current of $\frac{1}{2}NI$; right: Permeability as function of flux density for a standard steel for laminated yokes.}
\label{fig:resMag-permsteel}
\end{center}
\end{figure}

Let us consider an iron-dominated resistive magnet with a C-shaped yoke as shown in Fig.~\ref{fig:resMag-permsteel}; Ampère's law then becomes, with the integration path of the equation on the dotted line: $\oint_c \overrightarrow{H} \overrightarrow{{\rm d}l} = N I$.

When we take the average of the fields in the iron and in the pole gap, it follows that

\begin{equation}
N \cdot I = H_{\rm iron} \cdot l_{\rm iron} + H_{\rm air\,gap} \cdot  l_{\rm air\,gap}    \mathrm{ \hspace{1cm} thus \hspace{1cm}  } 
N \cdot I = \frac{B}{\mu_{0}\mu_{\rm r}}  \cdot l_{\rm iron} + \frac{B}{\mu_{0}} \cdot l_{\rm air\,gap}.  \label{eq:a5}
\end{equation}
As $\mu_{\rm r} \gg\mu_{0}$ in the iron, we get
\begin{equation}
N \cdot I =  \frac{B}{\mu_{0}}\cdot l_{\rm air\,gap}.  \label{eq:a6}
\end{equation}

For a magnet with a flux density of $1.5 \UT$ in a $50 \Umm$ pole gap we will need $NI = 59\,683 \UA$. This can be done with a coil of $2 \times 30$ turns, with $I = 994 \UA$. This implies a current density of $5 \UA/{\UmmZ}^{2}$ in a Cu conductor with a surface of $14 \times 14 \Umm^2$. Copper coils with air cooling can run with current densities of a few $\UAZ/\UmmZ^2$, while hollow water cooled copper conductors can be used up to a few tens of $\UAZ/\UmmZ^2$.
\par
\begin{figure}[h!]
\begin{center}
\includegraphics[height=5.5cm]{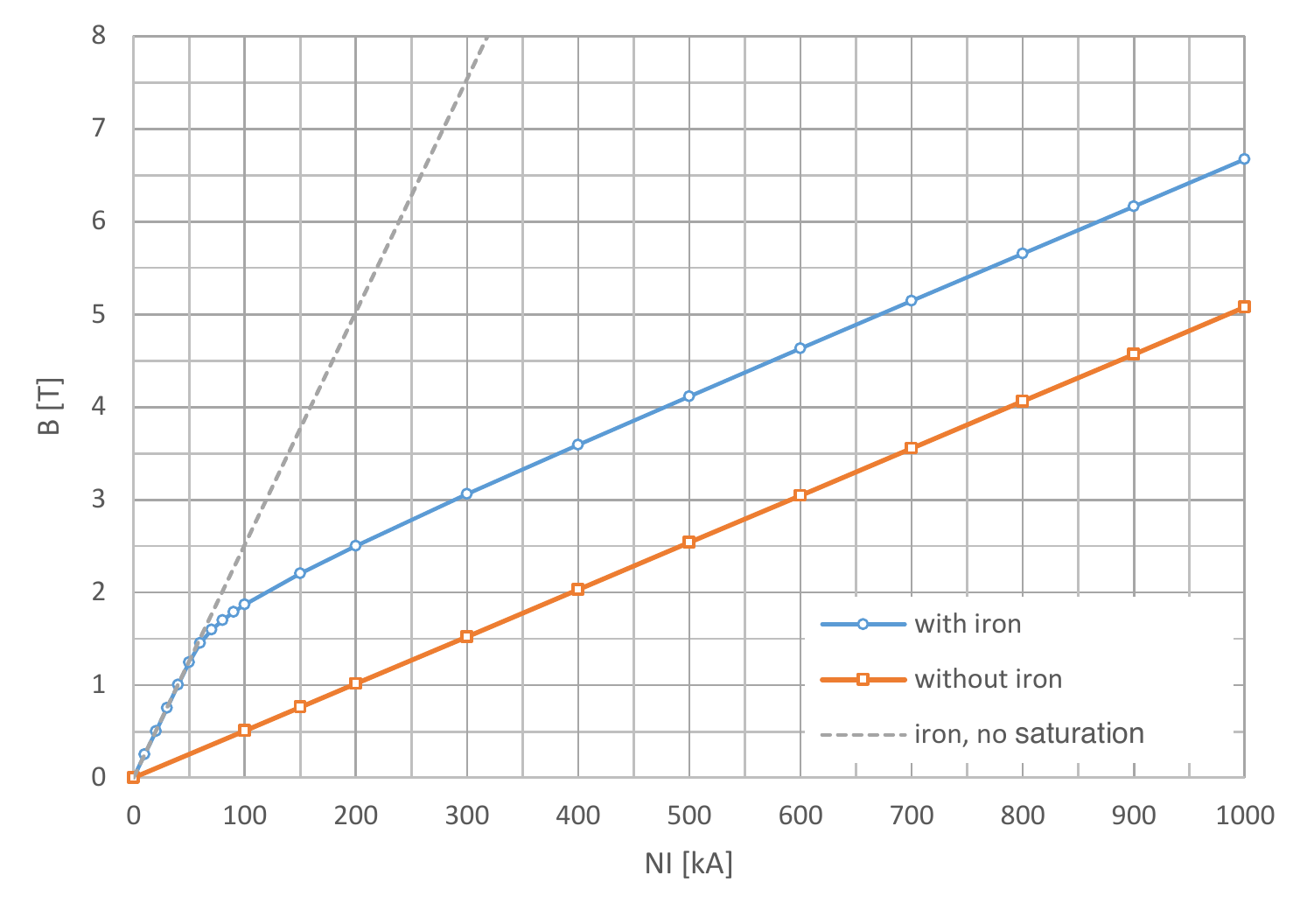}
\caption{The field of a magnet as function of the current for the cases with and without an iron yoke.}
\label{fig:airmagcomp}
\end{center}
\end{figure}

The permeability of the iron depends on the flux density and the best types of magnetic steel will be fully saturated close to $2\UT$.  In Fig.~\ref{fig:resMag-permsteel} the relative permeability $\mu_{\rm r}$ of a standard steel type used for laminated magnet yokes is shown. When the value of $\mu_{\rm r}$ approaches unity the approximation to derive~(\ref{eq:a6}) is not valid any more and we can see that the required current to reach a field $B > 2\UT$ will steeply increase. Iron-dominated magnets with $B > 2\UT$ very quickly become unpractical or impossible to build as the required current density in the coil exceeds the cooling possibilities.
In Fig.~\ref{fig:airmagcomp} we compare the field of a magnet with a vertical gap of $50 \Umm$ in the cases with and without an iron yoke.

\subsection{Magnetic field quality: Multipole description}
The beam in a particle accelerator is confined in a beam pipe around which the bending and focusing magnets have to be arranged.  From beam dynamics calculations one can derive that the field quality is very demanding; for the dipole magnets that provide the bending of the beam, the typical required field homogeneity is: $ \frac{\Delta B_{z}}{|B|} \leq 10^{-4} $. 
The field quality in accelerator magnets is expressed and measured in a multipole expansion:
\begin{equation}
B_{y}+{\rm i}B_{x} = 10^{-4}B_{1} \sum_{n=1}^{\infty}(b_{n}+{\rm i}a_{n}) \left ( \frac{x+{\rm i}y}{R_{\rm ref}} \right )^{n-1} \; \mathrm{with} \ b_{n}, a_{n} \leq \mathrm{few \ units}.  \label{eq:a8}
\end{equation}
With: $z=x+iy$; $B_{x}$ and $B_{y}$ the flux density components in the $x$ and $y$ direction; $R_{ref}$ the radius of the reference circle; $B_{1}$ the dipole field component on the reference circle; $b_{n}$ and $a_{n}$ the normal and skew $n^{th}$ multipole component. \\
In a circular accelerator, where the beam makes multiple passes, a typical demand is $a_{n}, b_{n} \leq 10^{-4}$.
\subsection{Magnetic length}
\begin{figure}[h!]
\begin{center}
\includegraphics[height=4.8cm, trim=0 2mm 0 4mm]{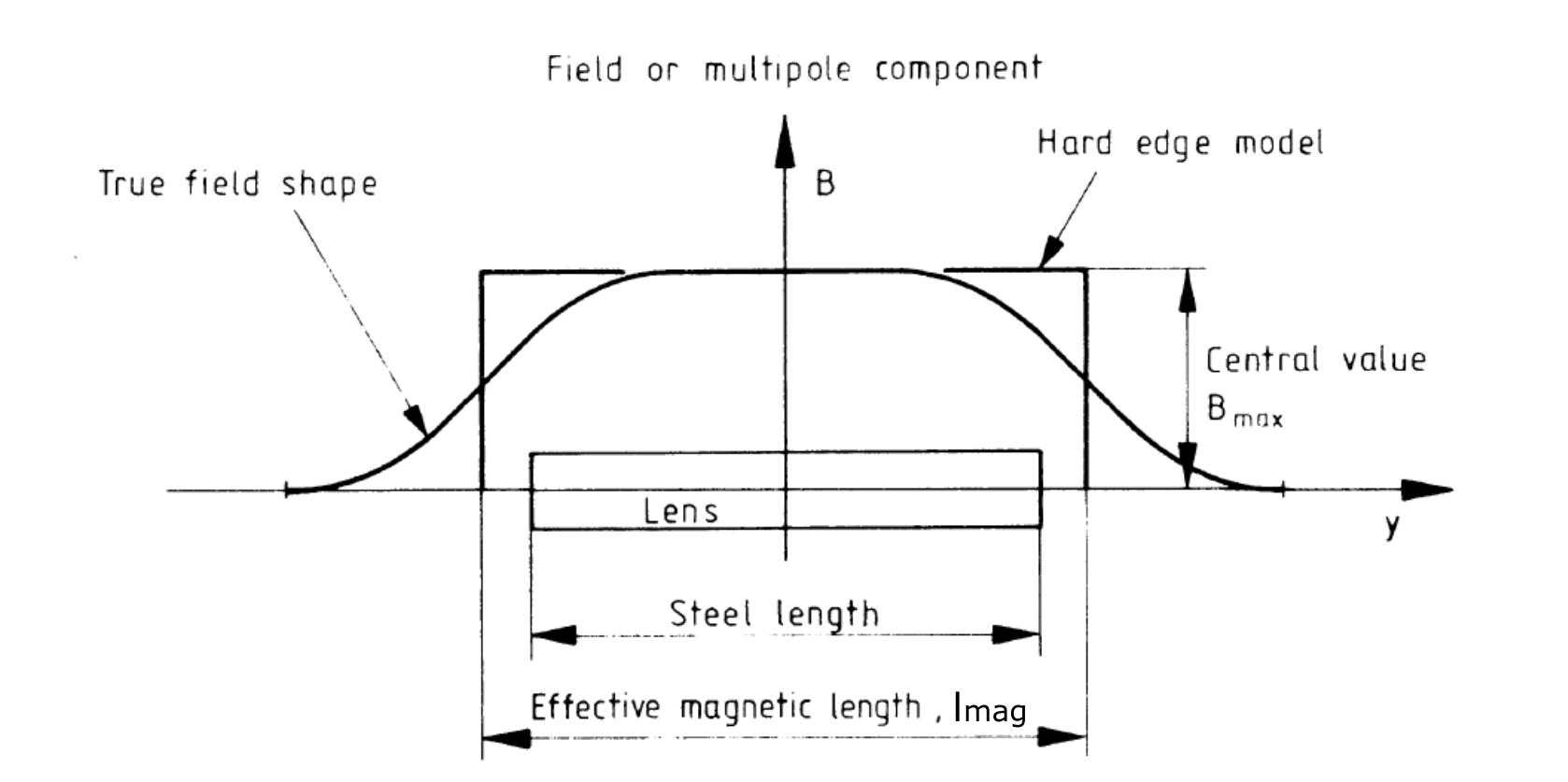}
\caption{Shape of the magnetic field in the longitudinal dimension (named y in this plot).}
\label{fig:axialfield}
\end{center}
\end{figure}

In the longitudinal dimension the typical shape of a magnetic field can be seen in Fig.~\ref{fig:axialfield}.
We can define the magnetic length $l_{mag}$ from: $ l_{mag}B_{0} = \int_{-\infty}^{+\infty}B(z)dz$ with $B_{0}$ the field at the centre of the magnet $z=0$. \\
Typically one can estimate that for warm dipole magnets $L_{mag}=L_{yoke}+d$ with $d$ the pole distance and for warm quadrupole magnets $L_{mag}=L_{yoke}+r$ with $r$ the inscribed radius between the poles.
\subsection{Magnets in an accelerator: powering circuits}
\begin{figure}[b!]
\begin{center}
\includegraphics[height=3.7cm, trim=0 2mm 0 2mm]{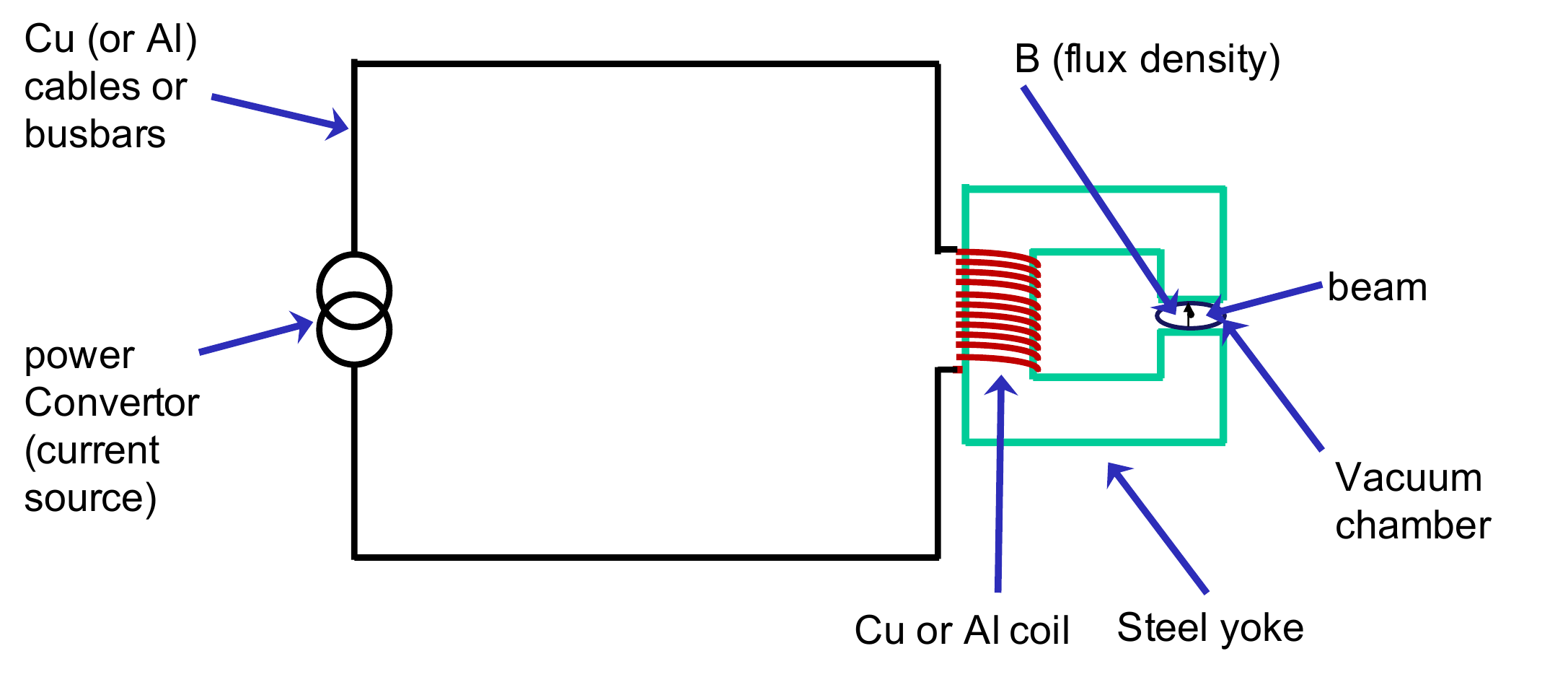}
\caption{Typical powering circuit for accelerator magnets.}
\label{fig:powercirc}
\end{center}
\end{figure}

We have to consider a magnet or a string of magnets in an accelerator as part of a powering circuit to be able to design the magnet with current and voltage values that are feasible for the magnet, the cabling and the power converter. A schematic drawing of a powering circuit can be found in Fig.~\ref{fig:powercirc}. 
We consider;  a typical $B$ field stability requirement in time in the order of $10^{-5}$ to $10^{-4}$; a current rise time during the acceleration cycle of $1\Us$ ; magnet resistance $R$ in the order of 20 m$\UOZ$ to 60 m$\UOZ$; magnet inductance $L$ in the order of 20 m$\UHZ$ to 200 m$\UHZ$. For the cables connecting the power converter to the magnet we take a length of $200\Um$ with a Cu cross section of $250\Umm^{2}$ for a current of $500\UA$ without cooling. With a specific resistance of Cu of 17 n$\UOZ\cdot\UmZ$ we get a resistance for the cables $R = 13$ m$\UOZ$. The power converter has to supply a current up to $500\UA$ with a stability of a few ppm and a voltage, for the resistive part of $40\UV$, and the inductive part (additional to the resistive part during ramp-up) of $100\UV$.

\subsection{Magnetic types}
\begin{figure*}[b!]
\begin{center}
\includegraphics[height=6.cm]{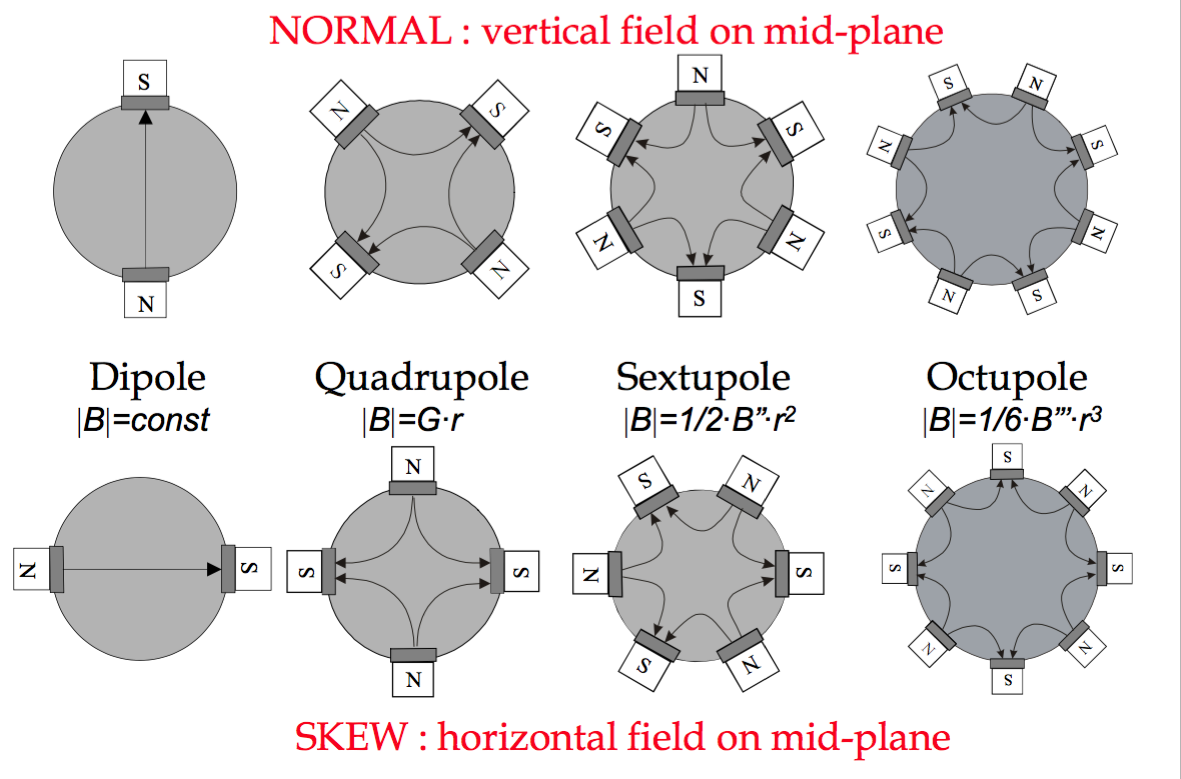} \hspace{ 2cm}
\includegraphics[width=10cm]{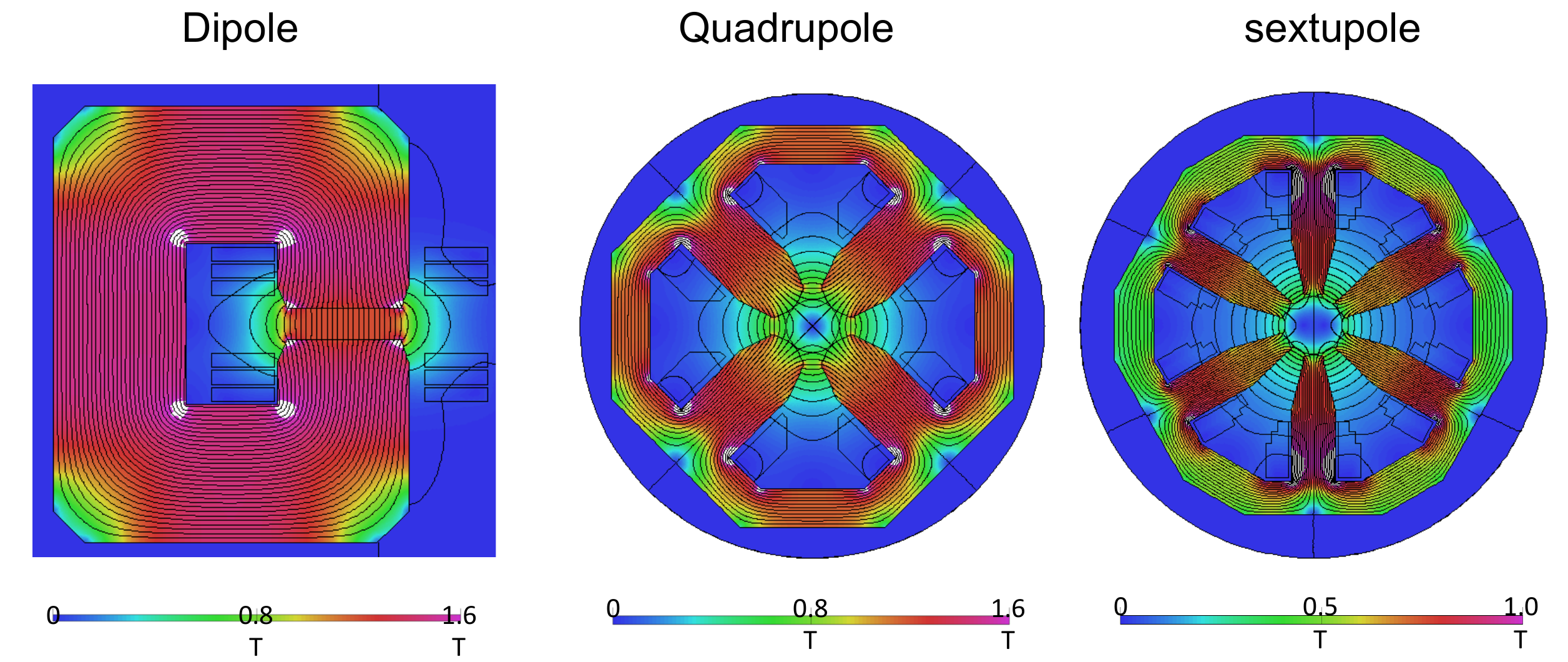}
\caption{Top: Magnets types according to the main field pole they supply up to $n=4$ for both normal and skew poles. Bottom: typical flux line patterns in dipole, quadrupole and sextupole magnets.}
\label{fig:normtypes}
\end{center}
\end{figure*}

In Fig.~\ref{fig:normtypes} the magnets types are shown according to the main field pole they supply up to $n=4$ for both normal and skew poles. One can also see the flux line patterns in a C type dipole a quadrupole and a sextupole.
Magnets are designed according to certain symmetries with mechanical errors also appearing with these symmetries. As a result, certain multipole harmonics can occur, the so-called "allowed harmonics" (see Table~\ref{tab:allowedharmon}).  
Non-symmetric designs and non-symmetric fabrication errors give rise to other, "non-allowed", harmonics.

\begin{table}[h!]
\begin{center}
\caption{Allowed harmonics for iron dominated magnets.}
\label{tab:allowedharmon}
\begin{tabular}{p{3cm}cc}
\hline\hline
\textbf{Magnet type}    & \textbf{Allowed harmonics $b_{n}$}  \\
\hline
$n=1$ Dipole              & $n=3,5,7, ...$        \\
$n=2$ Quadrupole     &  $n=6,10,14, ...$         \\
$n=3$ Sextupole        &  $n=9,15,21, ...$         \\
$n=4$ Octupole         &  $n=12,20,28, ...$         \\
\end{tabular}
\end{center}
\end{table}

\section{Practical magnet design and manufacturing}
To design and manufacture a magnet, we can identify the following steps:
\begin{enumerate}
\item{Specification;}
\item{Conceptual design;}
\item{Detailed design;}
	\begin{enumerate}
	\item{Yoke: yoke size, pole shape, FE model optimisation;}
	\item{Coils, cross section and cooling;}
	\item{Raw materials choice;}
	\item{Yoke ends, coil ends;}
	\end{enumerate}
\item{Yoke manufacturing, tolerances, alignment, structure;}
\item{Coil manufacturing, insulation, impregnation;}
\item{Magnetic field measurements.}
\end{enumerate}
In the following section we will go into more detail for each of these steps.
\subsection{Specification}
Before designing a magnet we have to get a precise specification from the accelerator designers. The specification should comprise:
\begin{itemize}
\item{Field or gradient.} \\ 
For a dipole the minimum and maximum flux density $B(\UTZ)$ should be specified. For a quadrupole this is the gradient $G(\UTZ/ \UmZ)$. For a sextupole this will be $G_{3}(\UTZ/ \UmZ^{2})$.
\item{Magnet type.} \\ 
For dipoles one has to choose between C-type, H-type, window, etc.. Further one has to know whether it will be a DC magnet, slow ramped or fast ramped (AC) magnet.
\item{Aperture.} \\ 
For dipole magnets one should define a rectangular "good field region" which then defines the air-gap height and width between the poles. Quadrupoles and other higher order magnets will have a circular "good field region" that determines the inscribed circle between the pole.
\item{Magnetic length and physical length.} \\ 
We need to know what the required magnetic length is and the allowed physical length. The coil will always stick out of the yoke and, depending on the coil end type selected, will take some space.
\item{Current range.} \\ 
The current range of the power converter needs to be specified and with it, the available voltage. 
\item{Field quality.} \\ 
The field quality in the "good field region" for a dipole $\frac{\Delta B}{B_{0}}$ and for a quadrupole $\frac{\Delta G}{G_{0}}$ need to be defined.
\item{Cooling type.} \\ 
Available cooling types are: air cooling, for which cooling surface and air flow rate are important, and water cooling, for which we need to know the available water pressure, allowed pressure drop and flow rate.
\item{Jacks and alignment features.} \\ 
The supporting systems of the magnet (jacks) have to be specified and how the alignment will be done (alignment targets).
\item{Vacuum chamber.} \\ 
The vacuum chamber and potential bake-out system will take space and need to be known at the beginning of the design.
\end{itemize}
These specification items will probably not be given as a complete list but will need negotiation with the various experts concerned and will probably be a compromise.
\subsection{Conceptual design}
I will give as an example a first step conceptual design of a C-type dipole magnet.

\begin{figure*}[h!]
\begin{center}
\includegraphics[height=4.4cm]{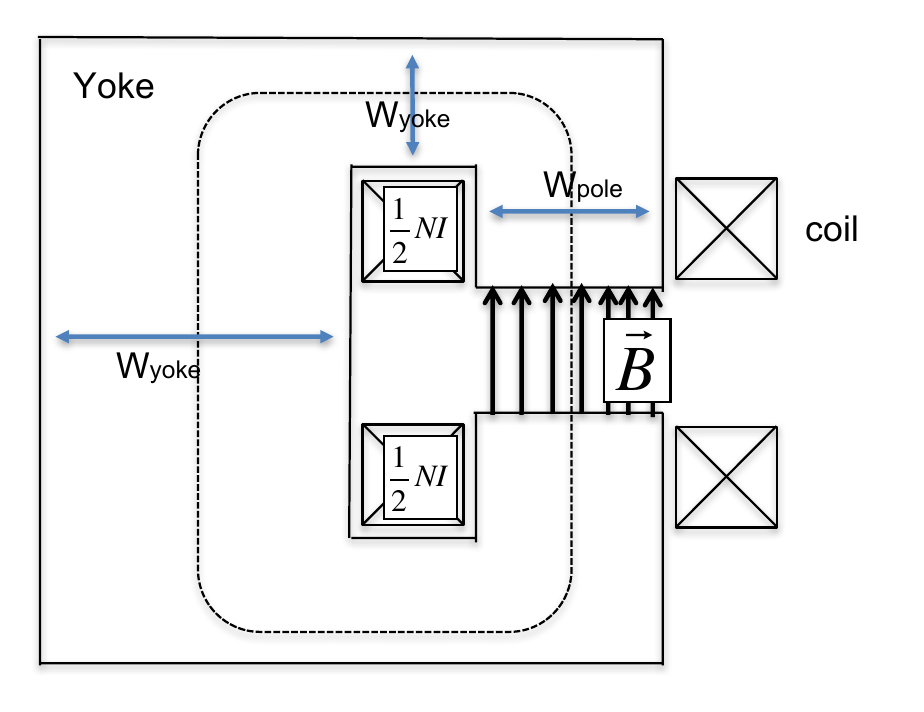} \hspace{0.cm}
\includegraphics[height=1.5cm]{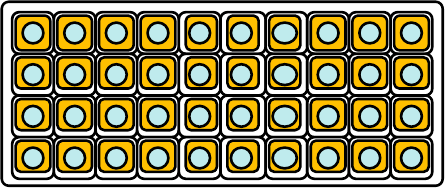}  \hspace{0.3cm}
\includegraphics[height=2.cm]{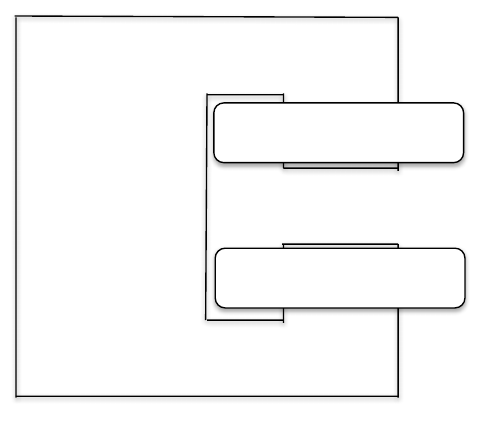}  \hspace{ 0.1cm}
\includegraphics[height=1.5cm]{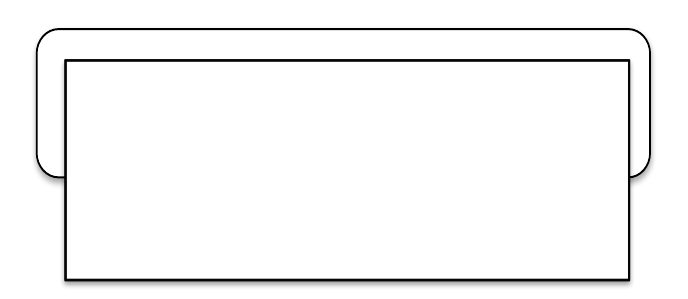} 
\caption{Left: Dipole cross section design, mid-left: coil cross section, mid-right: end view, right: top view.}
\label{fig:dipconsdesig}
\end{center}
\end{figure*}

\begin{itemize}
\item
With Equation~(\ref{eq:a6}) we can get $N \cdot I$ from $B$ and $l_{airgap}$ : $ N \cdot I = \frac{l_{airgap}  \cdot B} {\mu_{0}}$.
\item
We get from $ N \cdot I$ and the $I_{max}$ from the power converter the number of turns in the coil.
\item
We can then decide on the conductor cross section via the current density we will be using: $J_{coil} = 5 \UA/ \UmmZ^{2}$ for a water cooled coil or $J_{coil} = 1 \UA/ \UmmZ^{2}$ for an air-cooled coil.
\item
We can now define the coil cavity in the yoke. We need to allow for an insulation thickness of $0.5 \Umm$ around each conductor and a ground insulation around the coil of $1 \Umm$.
\item
With this we can now draw a first cross section of the magnet by estimating the required yoke thickness using the steel saturation flux density $B_{sat}$: $W_{yoke} = W_{pole} \frac{B}{B_{sat}}$. In general we will have $ 1.5\UT < B_{sat} < 2.0\UT$.
\item
Given the available longitudinal space we can decide on the geometry of the coil ends: racetrack or bedstead.
\item
We now have the rough magnet cross section and the outside envelope as shown in Fig.~\ref{fig:dipconsdesig} .
\end{itemize}

\subsection{Detailed design}
\subsubsection{Yoke}
The ideal poles for iron-dominated magnets are lines of constant scalar potential. For the first three orders they are:
\begin{itemize}
\item
Dipole:  \hspace{2cm} $y= \pm h/2$,  \hspace{2cm} a straight line,
\item
Quadrupole:  \hspace{1.2cm} $2 x y = \pm r^{2}$,  \hspace{1.85cm} a hyperbola,
\item
Sextupole:  \hspace{1.5cm}  $3 x^{2} y  - y^{3}= \pm r^{3}$.
\end{itemize}
To optimise the field homogeneity, small bumps, "shims", are added on the edges of the poles. \\
I will give an example for a dipole and for a quadrupole of this pole shimming.

\begin{description}
\item {Dipole:} \\
In Fig.~\ref{fig:lepdiplam} we can see a yoke lamination of the dipole of the LEP accelerator at CERN. On the edges of the poles we can see the pole shims.

\begin{figure*}[h!]
\begin{center}
\includegraphics[height=6.7cm]{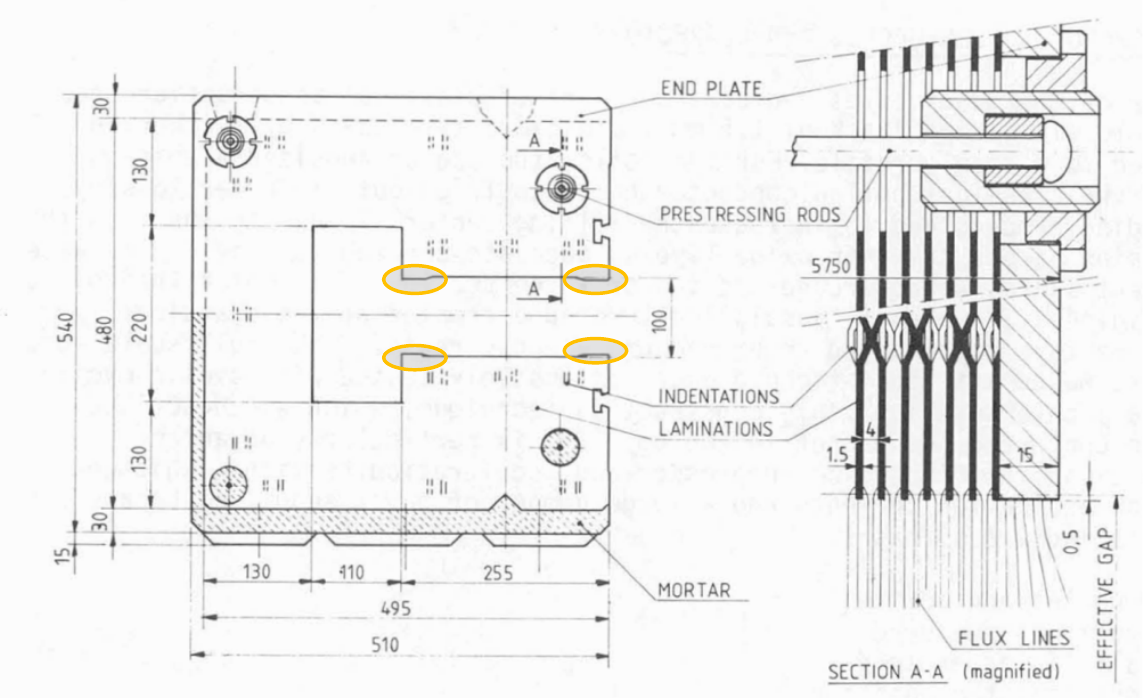}
\caption{C shaped yoke lamination from the (now dismounted) LEP accelerator at CERN. }
\label{fig:lepdiplam}
\end{center}
\end{figure*}

\item{Quadrupole:} \\
In the case of a quadrupole, very often also a small flat part that functions as alignment feature is added to the laminations to be able to assure a precise pole distance. In the left top image of Fig.~\ref{fig:quadlam} we can see this schematically indicated. In the right top image we can see the details of a pole with all features included. To actually measure the pole distance in a magnet purpose-build instrumentation can be used. In the figure we can see such a pole distance measurement device and the resulting measurement. 

\begin{figure*}[h!]
\begin{center}
\includegraphics[height=3cm]{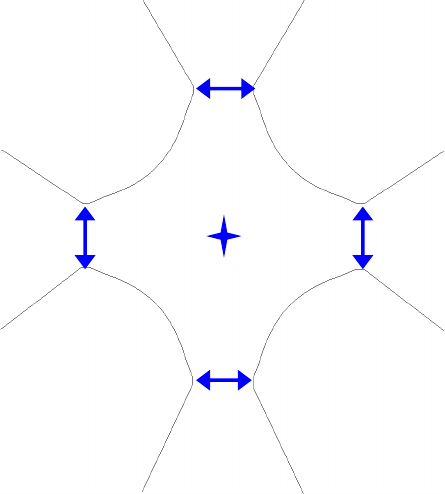} \hspace{0.5cm}
\includegraphics[height=2.9cm]{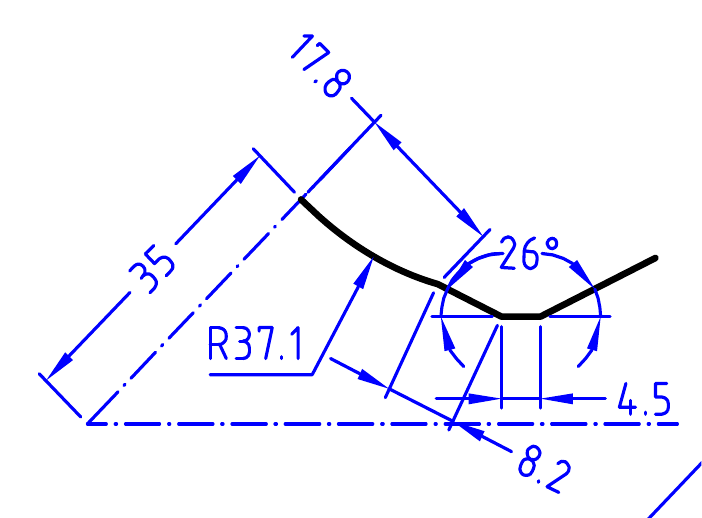}  \hspace{0.5cm} \\
\vspace{2mm}
\includegraphics[height=5.2cm]{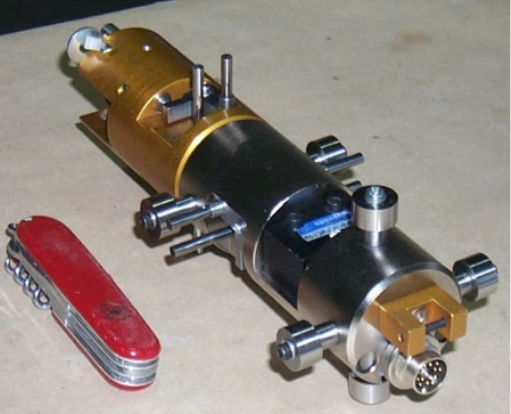}  \hspace{ 0.4cm}
\includegraphics[height=5.2cm]{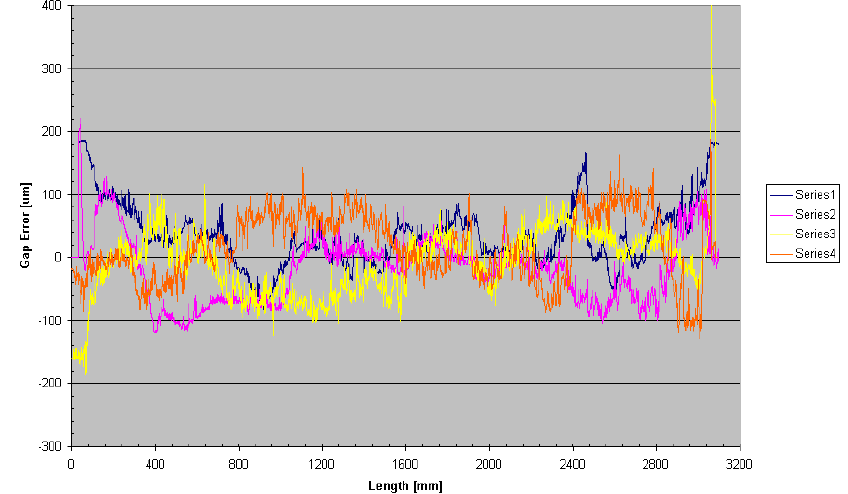} 
\caption{Quadrupole pole distance measurement.}
\label{fig:quadlam}
\end{center}
\end{figure*}

\end{description}

To design the detailed shape of the yoke, we need to use an electromagnetic (EM) Finite Element (FE) model. Using EM FE models, we can optimise the field quality by adjusting the pole shape and minimising high saturation zones in the yoke. We can then also minimise the total amount of steel to be used to reduce the magnet weight and cost. Finally the EM FE models are used to precisely predict the flux density magnitude ($B_{centre}$, $\int Bdl$, $ L_{mag}$) and the quality ($b_{n}$  and $a_{n}$). These values  are needed for the beam dynamics models in the accelerator design. 
Some EM FE codes employed are:
\begin{itemize}
\item "Opera" from Cobham: 2D and 3D EM FE commercial software \\
(see: http://operafea.com/ ),
\item "Poisson" 2D EM FE open software , now distributed by LANL-LAACG \\
(see: http://laacg.lanl.gov/laacg/services/download\_sf.phtml ),
\item "ROXIE" 2D and 3D EM BEM-FEM software, licensed by CERN \\
(see: https://espace.cern.ch/roxie/default.aspx ),
\item "ANSYS Maxwell" 2D and 3D commercial software \\
(see: http://www.ansys.com/Products/Electronics/ANSYS-Maxwell ).
\end{itemize}

\begin{figure*}[h!]
\begin{center}
\includegraphics[height=8.5cm]{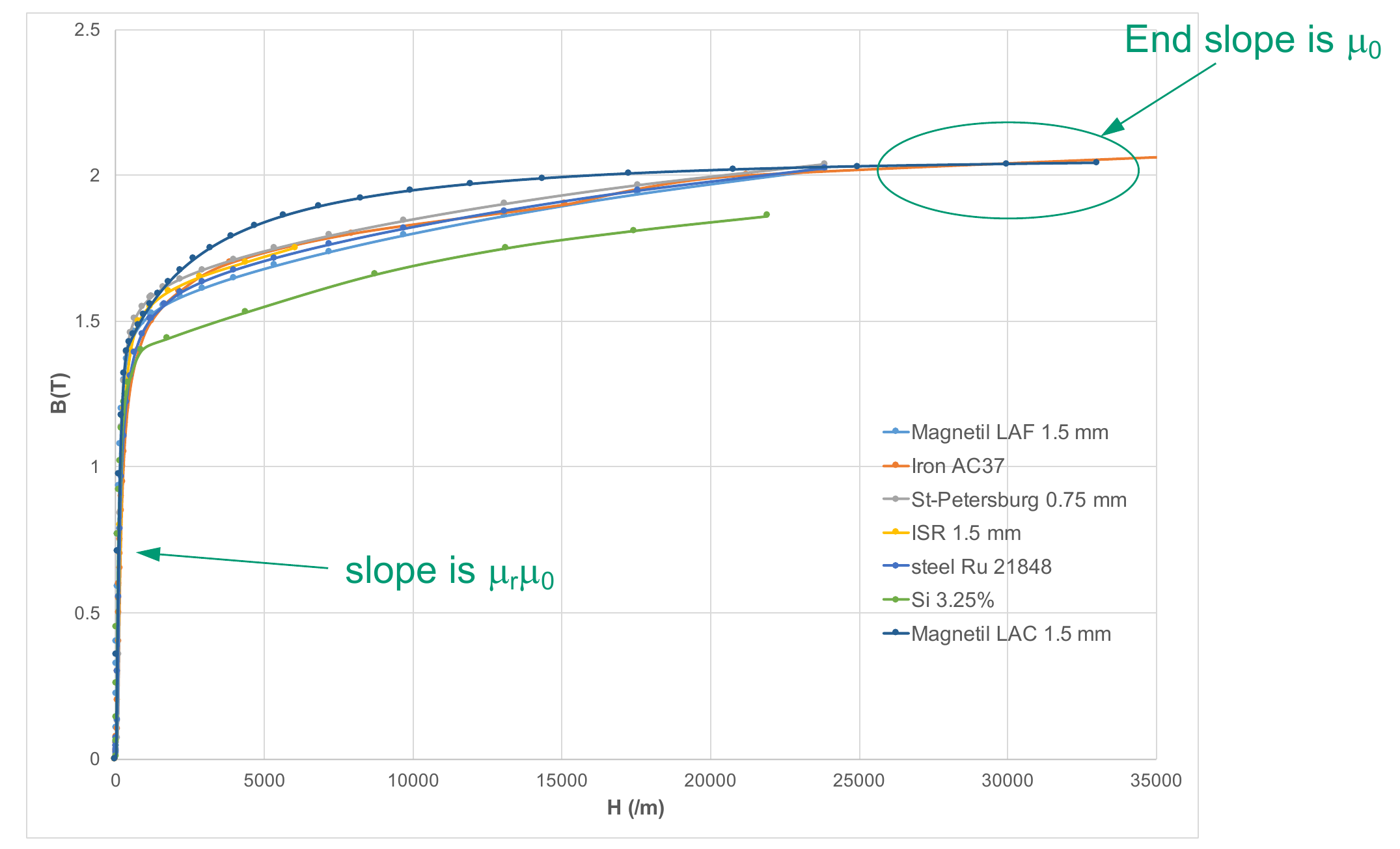}
\caption{ $B-H$ curves for various types of magnetic steel.}
\label{fig:satcurves}
\end{center}
\end{figure*}

\begin{figure*}[btp!]
\begin{center}
\includegraphics[height=5.5cm]{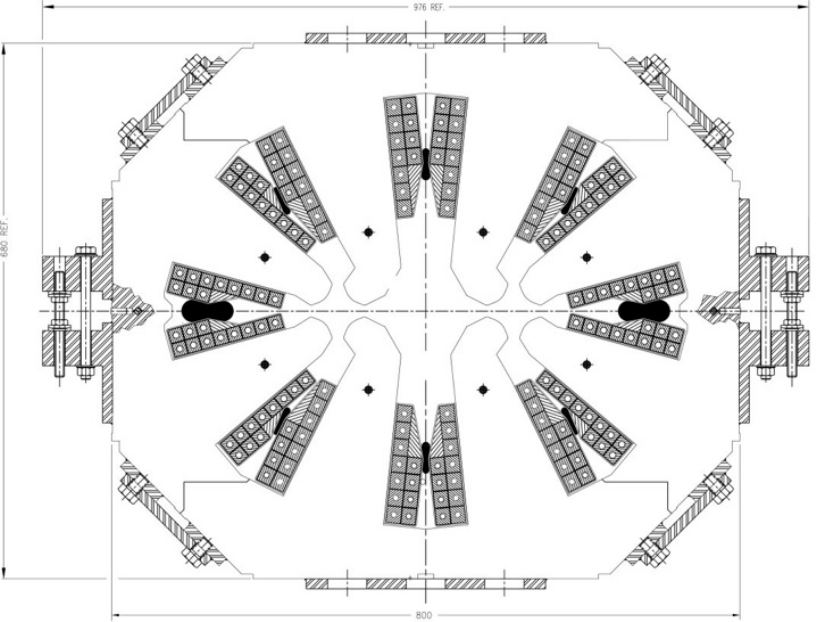}  
\includegraphics[height=5.cm]{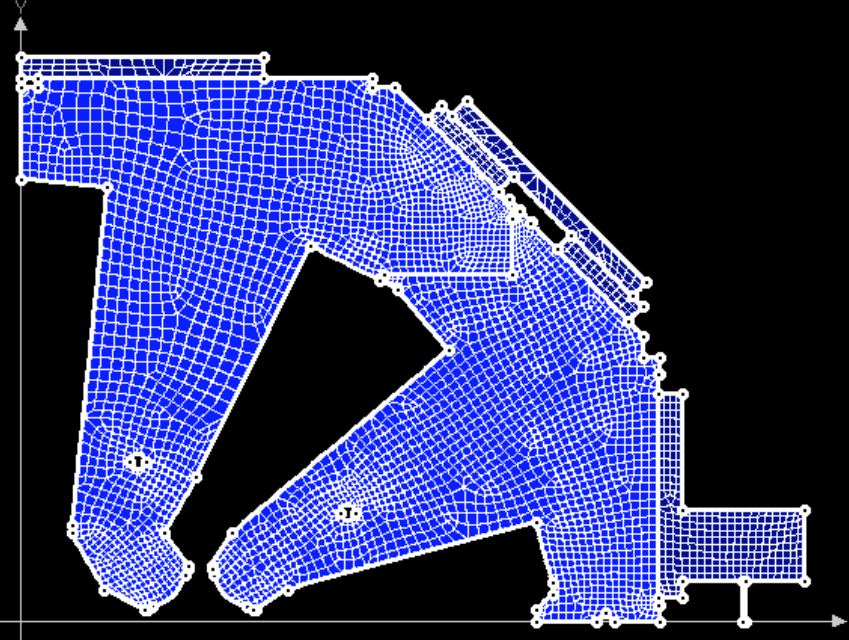} \\
\vspace{1mm}
\includegraphics[height=3.5cm]{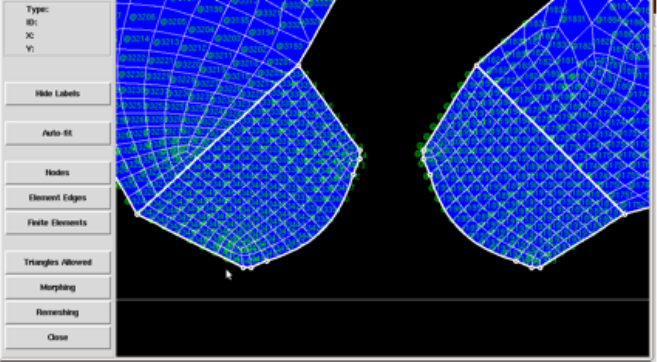} 
\includegraphics[height=5.5cm]{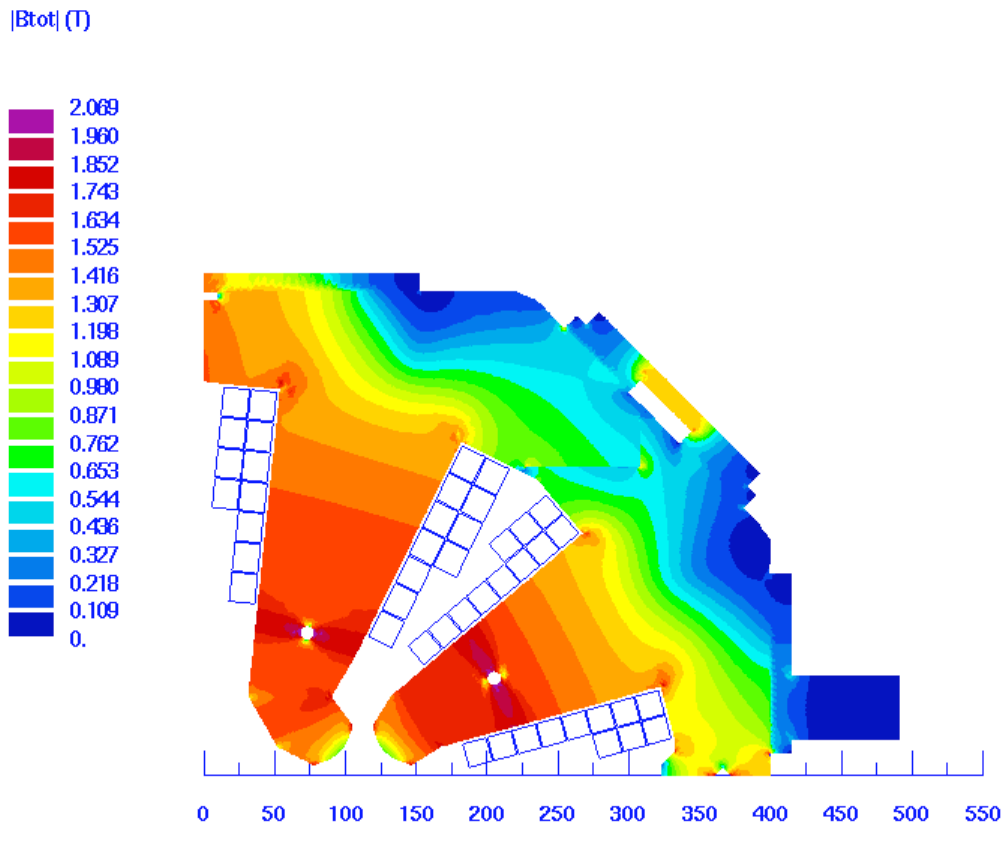} 
\caption{ROXIE EM model of the LHC MQW double aperture quadrupole. Top-left: cross section, top-right: mesh, bottom-left: mesh on poles, bottom-right: calculated nominal field ($G=35\UT/\UmZ$ at $I=710\UA$).}
\label{fig:EMFEexample}
\end{center}
\end{figure*}

The EM FE models require as input a description of the $B(H)$ relation of the yoke material. To get the correct final field values, we have to use a measured, and smoothed, curve of the material used for the yoke. In Fig.~\ref{fig:satcurves}, a number of examples can be found of $B(H)$ curves for magnetic steel types. Fig.~\ref{fig:EMFEexample} shows an example of a ROXIE model for a double-aperture quadrupole (MQW) in the LHC at CERN.

\subsubsection{Coils}
\begin{figure*}[tbp!]
\begin{center}
\includegraphics[height=4.2cm]{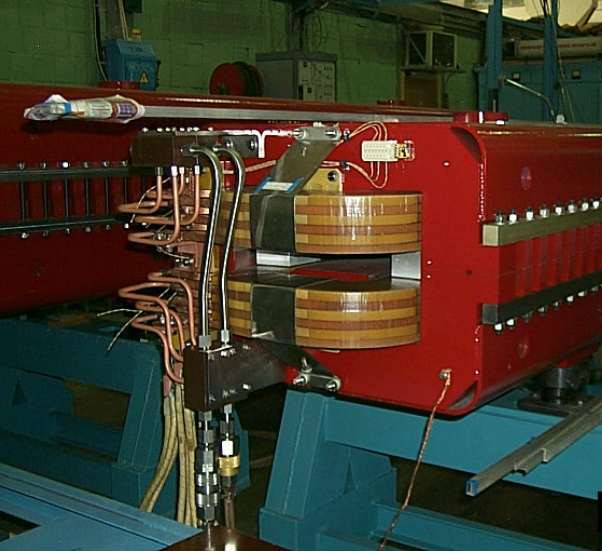}  \hspace{0.5cm}
\includegraphics[height=4.2cm]{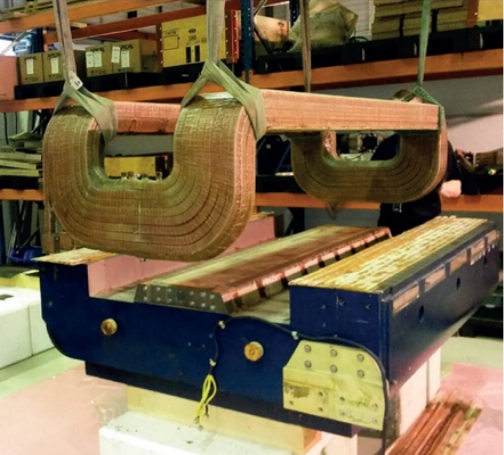} \hspace{0.5cm} \\
\vspace{2mm}
\includegraphics[height=4.2cm]{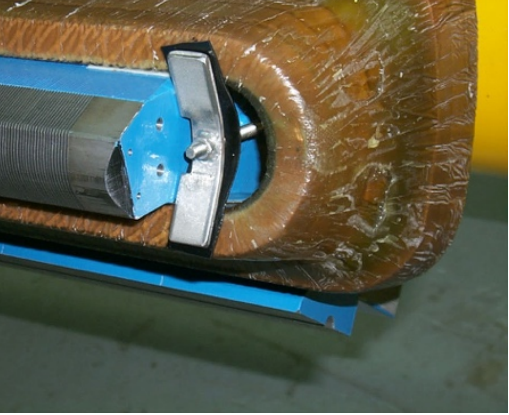} \hspace{0.5cm}
\includegraphics[height=4.2cm]{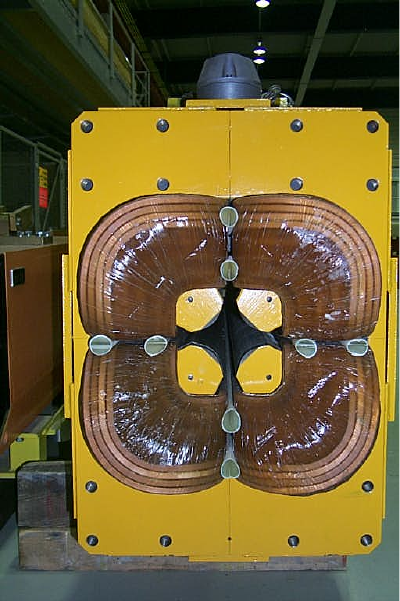} 
\caption{Top-left: dipole racetrack coil end, top-right: dipole bedstead coil end, bottom-left: quadrupole racetrack coil end, bottom-right: quadrupole bedstead coil end.}
\label{fig:coilends}
\end{center}
\end{figure*}

The straight part of the coils are of rather straightforward geometry but the end of the coils will need some attention in the design. Several types of coil ends are available but in practice they are variations of the two main types: racetrack and bedstead.  In Fig.~\ref{fig:coilends} the two types for dipoles and quadrupoles are illustrated.

There are essentially two types of coils: air cooled and water cooled. In the case of air cooling, the conductor is an insulated copper wire and the heat is passing by conduction to the outer surface of the coil where the air takes the heat away. It is generally applied to rather small coils for corrector magnets. Water cooled coils are applied for both small and large magnets working at higher current density. The conductors are hollow copper or aluminium bars surrounded by insulation with the water circulating in the hole where the heat is removed locally. \\
Practically all accelerator magnet coils are impregnated with epoxy to provide a good electrical insulation. Depending on the radiation load by beam residues, this epoxy might need to be radiation resistant. For magnets with extremely high radiation loads, other solution than epoxy have to be employed. Already in the 1970s completely mineral solutions were employed for magnets close to the SPS primary targets on quadrupoles featuring concrete as insulation material.  \\

The current in the resistive material of the coil will generate power that will heat up the coil. Let us first look at the power generated in the coil.
At a constant current (DC case), we can calculate the power from the length of conductor in a coil $N\cdot L_{turns}$, the conductor cross section $\sigma$, and the specific resistivity of the material $\rho$:
\begin{equation}
 \frac{P}{l} = \frac{\rho}{\sigma} I^{2} \hspace{0.5cm} 
 \end{equation}
with:	
\begin{equation}
\rho_{Cu}= 1.72(1+0.0039(T-20))10^{-8}\UO \UmZ  \hspace{0.5cm} or 	\hspace{0.5cm} 	\rho_{Al}= 2.65(1+0.0039(T-20))10^{-8}\UO \UmZ . 
\label{eq:a12}
\end{equation}
In the case of pulsed magnets we take the average of $I^{2}$ for the duty cycle. \\
The coil will have most of the power losses although also the yoke will experience losses due to hysteresis and eddy currents. 
The power losses due to the hysteresis in the yoke can be taken as (Steinmetz law up to $1.5 \UT$): 
\begin{equation}
 P[\UWZ / \UkgZ] = \eta f B^{1.6}  \hspace{0.2cm}  \textrm{with}  \hspace{0.2cm} \eta = 0.01 \hspace{0.1cm}  \textrm{to}  \hspace{0.1cm}  0.1 , \hspace{0.2cm} \eta_{Si steel} \approx 0.02 ,  \hspace{0.2cm} f  \textrm{[Hz]},  \hspace{0.2cm} B  \textrm{[T]}
 \end{equation}
 Eddy currents in the yoke will cause power losses as:
 \begin{equation}
 P[\UWZ / \UkgZ] = 0.05( d_{lam} \frac{f}{10} B_{av})^{2}
 \end{equation}
with $d_{lam}$ the yoke lamination thickness in mm and $B_{av}$ the average flux density at top field in the yoke in T.  

The parameters of water cooling that are to be defined are: $d_{cooling}$: the diameter of the cooling hole in the hollow conductor; $P_{water}$ [bar]; $\Delta P$ [bar] : the pressure drop over the coil; $Q$ [l/min]: the flow rate. We then: 
\begin{itemize}
\item
Choose the desired $\Delta T$. ( $20 \UDC$ or  $30 \UDC$ depending on the $T_{cooling water}$).
\item
With the heat capacity of the water ($4.186 \UkJ/ \UkgZ \UDCZ)$ we now know the required water flow rate $Q$[l/min].
\item
The cooling water needs to be in a moderate turbulent regime to optimise the heat transfer (laminar flow has zero flow speed on the wall) : Reynolds > 2000.
\begin{equation}
     R_{e} = \frac{d \cdot v}{\nu} \approx 140 d [\UmmZ]v \hspace{0.5cm}  \textrm{with }  
\end{equation}
\begin{equation}
     v  \textrm{ water velocity [m/s], } \nu  \textrm{ dynamic viscosity}  \approx \textrm{0.7x10$^{-6}$ m}^{2} \textrm{/s for } T_{water} \approx 40  \UDC  
 \end{equation}
\item
Get a good approximation for the pressure drop in smooth pipes from the Blasius law:
\begin{equation}
\Delta P  \textrm{[bar]} = 60l [ \UmZ ] \frac{Q  \textrm{[l / min]} ^{1.75} } {d [ \UmmZ ] ^{4.75}} 
 \end{equation}
\end{itemize}

\subsubsection{Raw material choice}
Magnet yokes are nearly always laminated to reduce the eddy currents during current ramping. Therefore, the yoke raw material has be thin steel sheets from which the laminations can be cut or punched. The magnetic properties of the steel depend on chemical composition, temperature treatments and the mechanical history. Important parameters for the magnetic performance of the steel are the coercive field $H_{c}$ and the saturation flux density $B_{sat}$.  Low carbon steel (with a carbon content $< 0.006\%$) is best suited for higher fields $B > 1 \UT$. The coercive field $H_{c}$ has an impact on the remnant field at low excitation currents, which is important for magnets that are also used at very low field for beam injection. Typically the specification is $H_{c} < 80 \UA / \UmZ$, and for magnets ranging down into low fields $B < 0.05 \UT$ this becomes $H_{c} < 20 \UA / \UmZ$.
Laminations can be coated with an inorganic (oxidation, phosphating, Carlite) or organic (epoxy) layer to increase the inter-lamination resistance. These coated steel sheets can be supplied by the steel manufacturers. In Tables~\ref{tab:steel} and \ref{tab:steel2} the properties of two types of steel can can be found.

\begin{table}[h]
\begin{center}
\caption{Specification for Low carbon magnetic steel, oxide coated, LHC separation dipoles (high field), 1.5 mm thick.}
\label{tab:steel}
\begin{tabular}{p{3cm}cc}
\hline\hline
\textbf{H(A/m)}    & \textbf{B(T)  }\\
\hline
40             & 0.20       	\\
60		& 0.50       	\\
120       	& 0.95       	\\
500		& 1.4         	\\
1200		& 1.5        		\\
2500		& 1.62         	\\
5000		& 1.71         	\\
10000	& 1.81        	\\
24000	& 2.00       	\\
\end{tabular}
\end{center}
\end{table}

\begin{table}[h]
\begin{center}
\caption{Specification for Silicon magnetic steel, epoxy coated, LHC transfer line corrector dipoles (low field), 0.5~mm thick.}
\label{tab:steel2}
\begin{tabular}{p{3cm}cc}
\hline\hline
\textbf{H(A/m)}    & \textbf{B(T)  }\\
\hline
100		& 0.07       	\\
300       	& 1.05       	\\
500		& 1.35         	\\
1000		& 1.5        		\\
2500		& 1.62         	\\
5000		& 1.72         	\\
10000	& 1.82        	\\
\end{tabular}
\end{center}
\end{table}

Low resistance copper conductor for the coils can be procured from several suppliers in many formats. For water cooled magnets one can buy hollow Cu conductor in tens of meters of piece-length. Both copper and and aluminium enamelled wire, which is coated with a very thin layer of insulation, can be directly ordered from suppliers. This insulation is often a tough polymer film. 

\subsubsection{Yoke manufacturing}
\begin{figure*}[h!]
\begin{center}
\includegraphics[height=5.4cm]{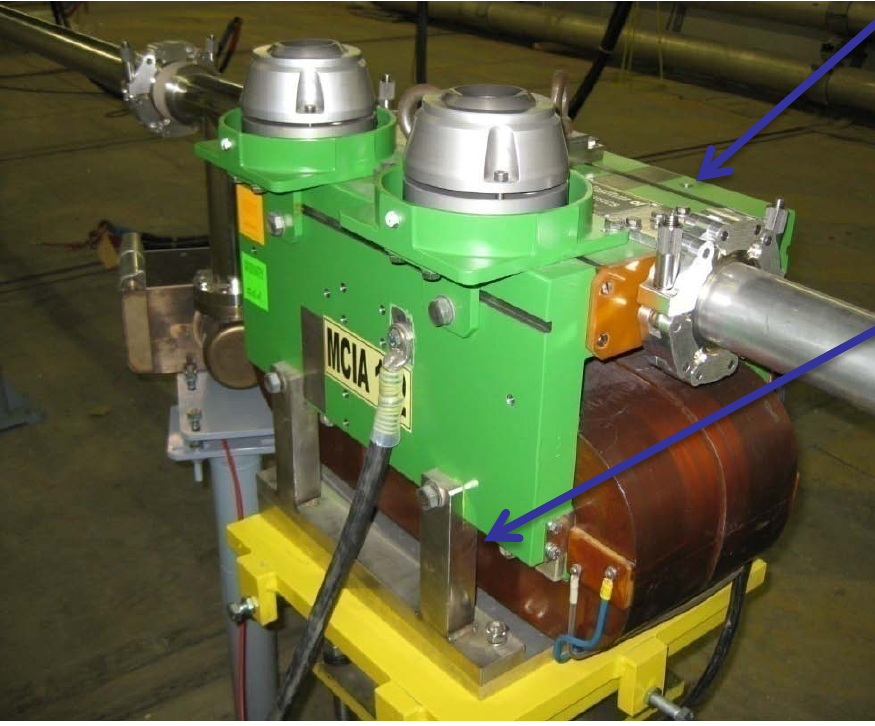} 
\includegraphics[height=5.4cm]{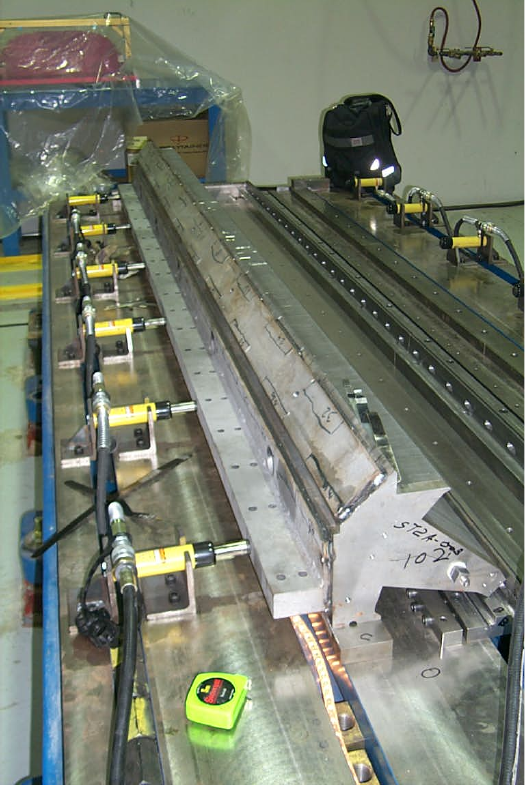}  
\includegraphics[height=3.7cm]{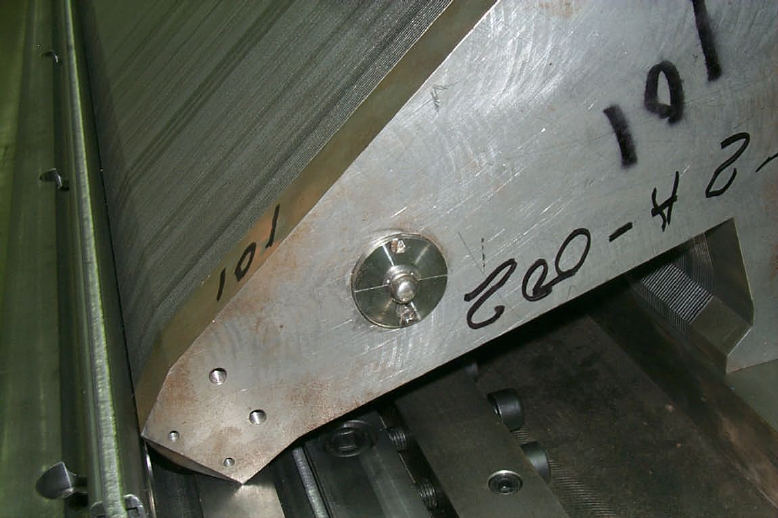} 
\caption{left: glued yoke of a small corrector magnet; middle: welded yoke stack of a 3 m long quadrupole magnet; right: tie-rod in the pole of a quadrupole yoke stack.}
\label{fig:yokefix}
\end{center}
\end{figure*}

The yoke of most warm magnets needs to be split into yoke-stacks so that the finished coils can be put around the poles. For an "H-shaped" vertical field dipole, we need to split the yoke at the horizontal plane in 2 yoke-stacks or half yokes. Quadrupole yokes habitually need to be split into 4 yoke-stacks. A "C-shaped" dipole yoke does not have to be split if the pole distance is sufficiently large to allow for sliding in the coils.
In general, yoke-stacks are either glued, using epoxy coated laminations, welded, with the support of full length structural plates on the outside or axially compressed by tie-rods (Fig.~\ref{fig:yokefix}). Often a combination of 2 or 3 of these are used. To stabilise the yoke-stack, thick steel end plates are used that can be either a glued stack of laminations or a solid steel plate. The end plates can have a chamfer on the pole to shape the end field. \\

\begin{figure*}[h!]
\begin{center}
\includegraphics[height=4.3cm]{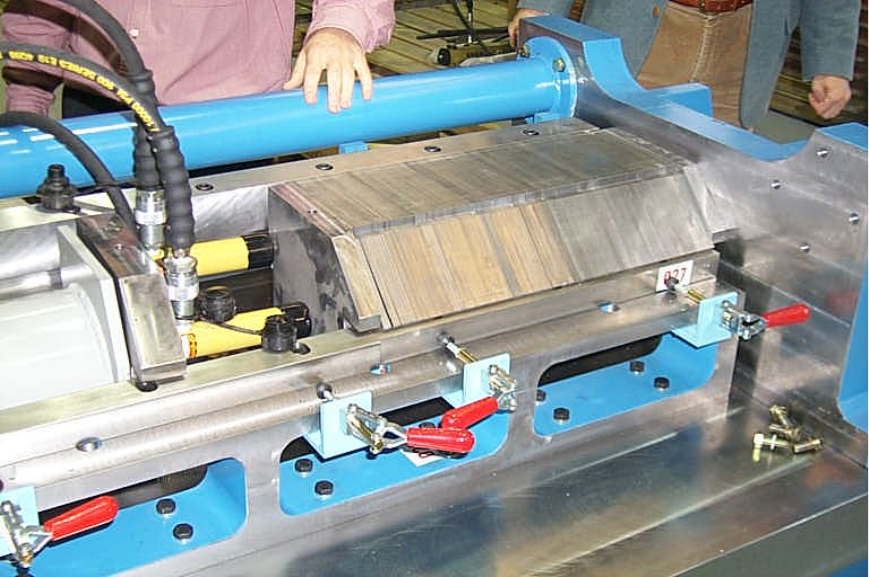} 
\includegraphics[height=4.3cm]{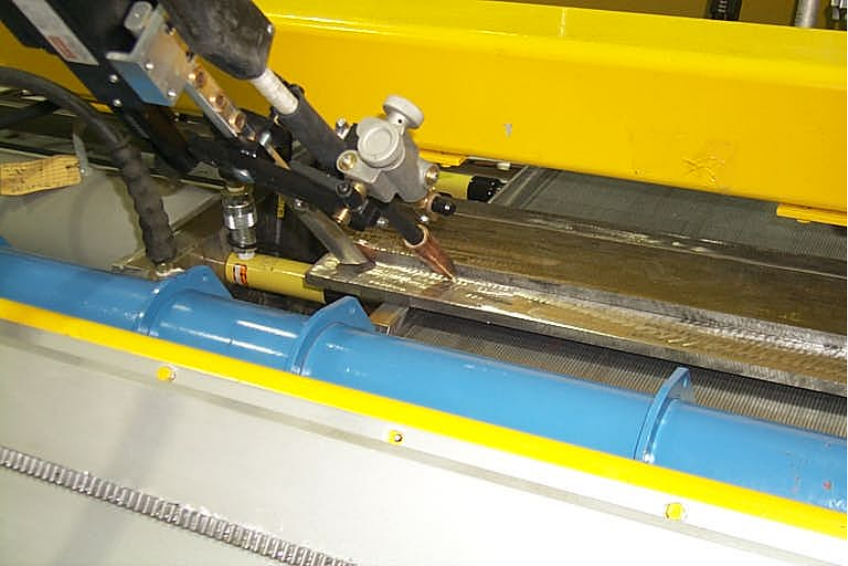}  
\includegraphics[height=4.3cm]{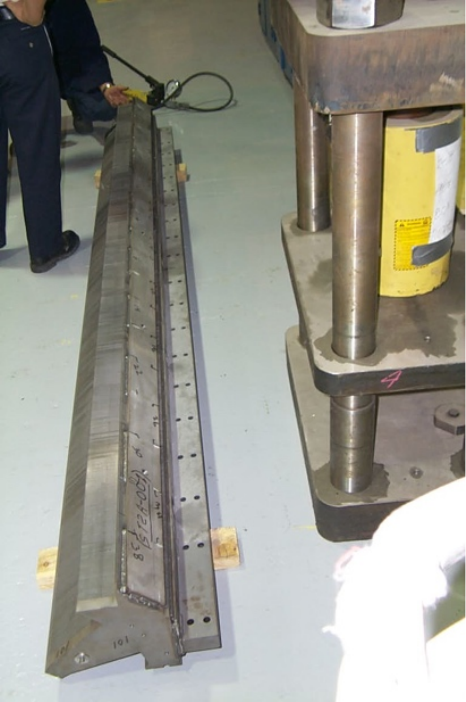} 
\caption{left: yoke-stack stacking; middle: yoke-stack welding; right: finished yoke stack before painting.}
\label{fig:stackmanuf}
\end{center}
\end{figure*}

The productions steps of a magnet yoke-stack are (Fig.~\ref{fig:stackmanuf}).
\begin{itemize}
\item
Lamination cutting or punching.
\item
End plate machining.
\item
Yoke-stack stacking.
\item
Yoke stack welding or gluing (or both).
\item
Mount tie rods to the yoke stack.
\item
Geometry verification measurements (straightness, flatness, etc).
\item
Cleaning and painting.
\end{itemize}
 
\pagebreak

\subsubsection{Coil manufacturing}
The steps for fabricating a coil with internal water cooling are (Fig.~\ref{fig:coilmanuf}):

\begin{figure*}[b!]
\begin{center}
\includegraphics[height=3.6cm]{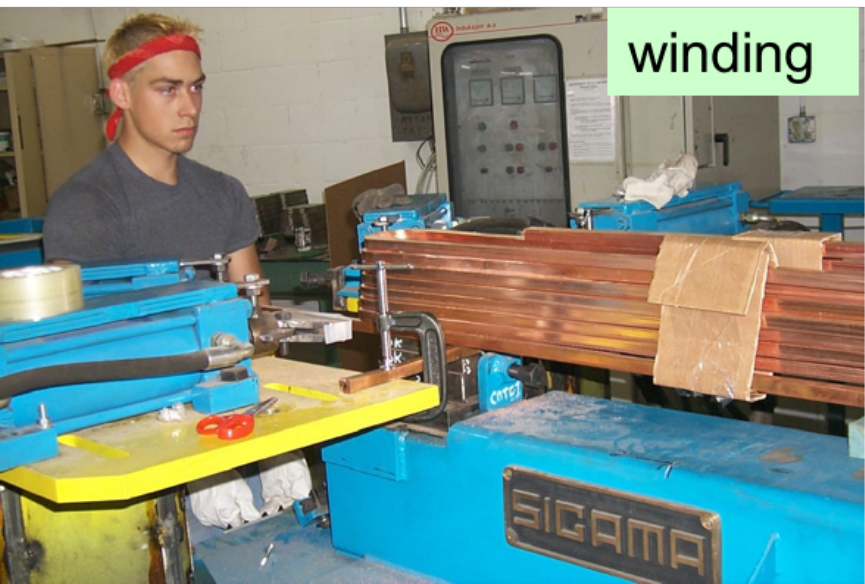}  
\includegraphics[height=3.6cm]{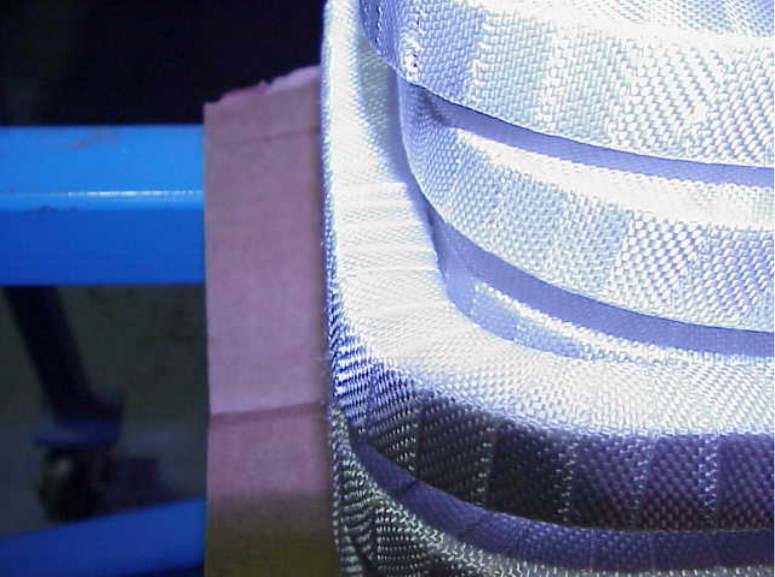}  
\includegraphics[height=3.6cm]{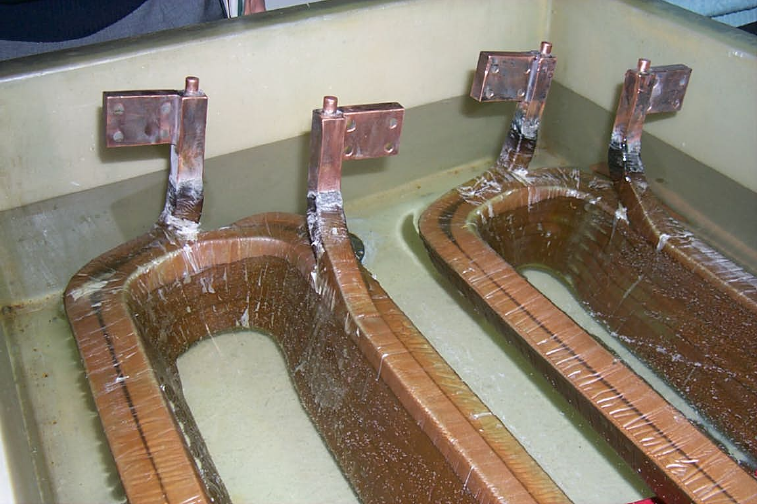} 
\caption{left: coil winding; middle: coil insulating; right: coil electrical testing in water.}
\label{fig:coilmanuf}
\end{center}
\end{figure*}

\begin{itemize}
\item
Hollow Cu conductor. \\
The conductor should be delivered in a format that will be directly mountable on the winding machine without further manipulations. As the Cu has to be soft ("dead soft"), it has to undergo a thermal treatment at the supplier ($\approx400\UDC$). Copper will harden when cold worked, so between the thermal treatment and the winding the copper should not be bend to allow for a smooth, low force winding.
\item
Glass fibre insulation. \\
The individual conductor has to be wrapped half-lapped with a glass-fibre tape with a thickness of $0.25 \Umm$ giving a total thickness of $0.50 \Umm$ to get a full coverage with no holes and assured spacing between the turns. The wrapping can be done on the winding machine just in front of the coil (for thin conductors requiring a small force to wind) or after winding by spreading out the turns and then wrapping the conductor by hand (for thicker conductors requiring a large winding force that then can damage the glass-fibre).
\item
Winding the conductor into a coil. \\
This operation is done on special winding machines. Depending on the type of coil, such machines are commercially available or are purpose built.
\item
Mount connections. \\
Hard solder the electrical connecting flags and the water connections to the coil leads.
\item
Apply ground insulation. \\
The coil as a whole has to be wrapped half-lapped with a glass-fibre tape with a thickness of $0.50 \Umm$ giving a total thickness of $1.00 \Umm$. 
\item
Mount the coil in an impregnation mould and impregnate with epoxy.
\item
Electrical tests. \\
The electrical turn-to-turn insulation should be verified. Large defects can be found by measuring the electrical resistance with a precision better than the resistance of one turn. To detect smaller insulation inter-turn defects a capacitive discharge method can be used. The ground insulation of the coil should be tested by immersion of the coil into water (with the connections sticking out) for several hours after which the resistance between the bath and the conductor should be measured. Depending of the required operational maximum voltage of the magnet in the accelerator (turn-to-turn in the coil, and coil to ground) electrical engineering standards will prescribe the test voltages (e.g. a factor 2 + $500 \UV$) that the coil has to be able to withstand. 
\end{itemize}

\subsubsection{Magnet assembly}
\begin{figure*}[h!]
\begin{center}
\includegraphics[height=5cm,trim=0 0.1cm 0 0]{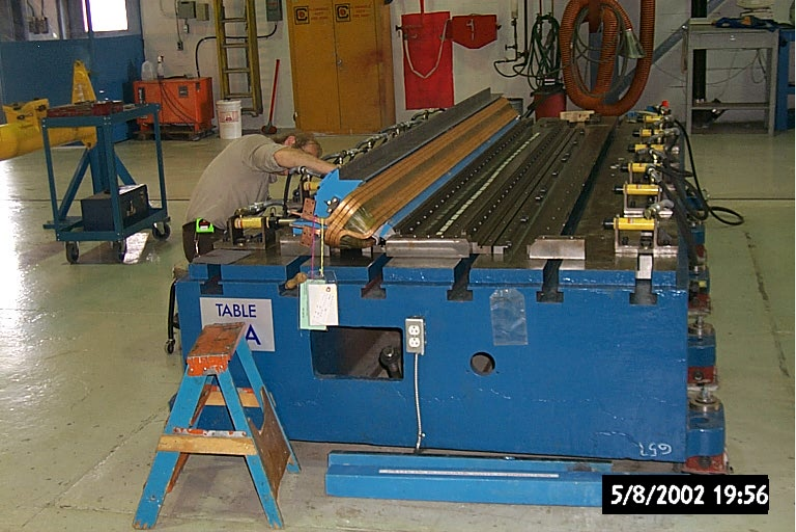}\hspace{6mm}
\includegraphics[height=5cm,trim=0 0.1cm 0 0]{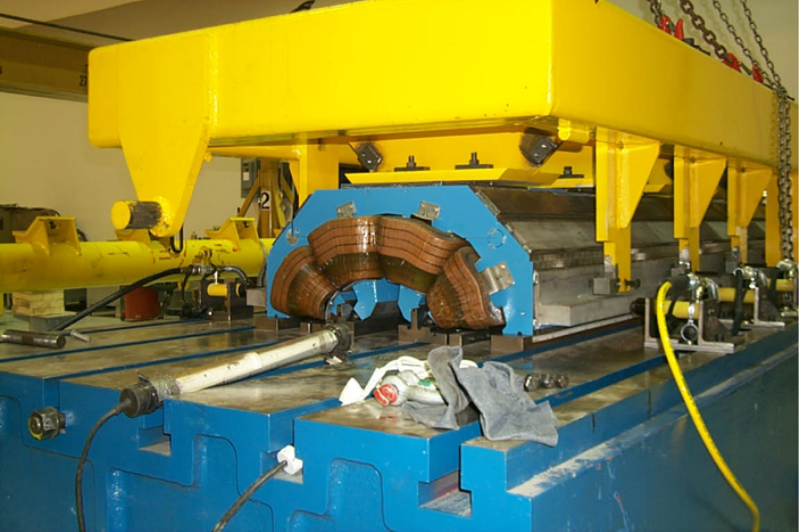} 
\caption{Left: yoke stack on a precision assembly table; right:half magnet handling tooling.}
\label{fig:assem}
\end{center}
\end{figure*}

When the yoke-stacks and all the coils are finished the magnet can be assembled. Typical steps for the magnet assembly are (Fig.~\ref{fig:assem}):
\begin{itemize}
\item
Mounting a coil on each pole on the yoke stacks. 
\item
Assemble the yoke-stacks. \\
Sometimes an intermediated step is done by first making half magnets and then joining the two together. Dedicated lifting girders are needed for these operations. For the magnet mounting, special tooling is needed to assure the proper alignment (e.g. assembly tables or jigs).
\item
Connection of the coils. \\
Once the yoke-stacks with coils are assembled, the electrical connections have to be made. This can be done by hard-soldering copper bars between the connection lugs. The water connections are then also to be made. Water cooling can be done for the coils in parallel or in series depending on the required cooling conditions. 
\item
Add instrumentation. \\
On water cooled magnets, temperature sensors or switches are mounted on each coil to be able to switch the current off in case of over-heating.
\item
Dimensional measurements. \\
The straitness of the magnet and the pole distance as function of the longitudinal position have to be measured.
\end{itemize}
\vspace*{-7.5mm}
\subsubsection{Magnetic field measurements}
\begin{figure}[pt!]
\begin{center}
\includegraphics[height=4.5cm]{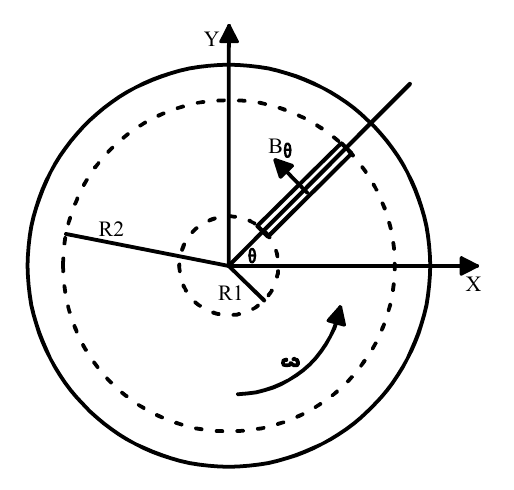}  
\includegraphics[height=5.cm,trim=0 0.3cm 0 0]{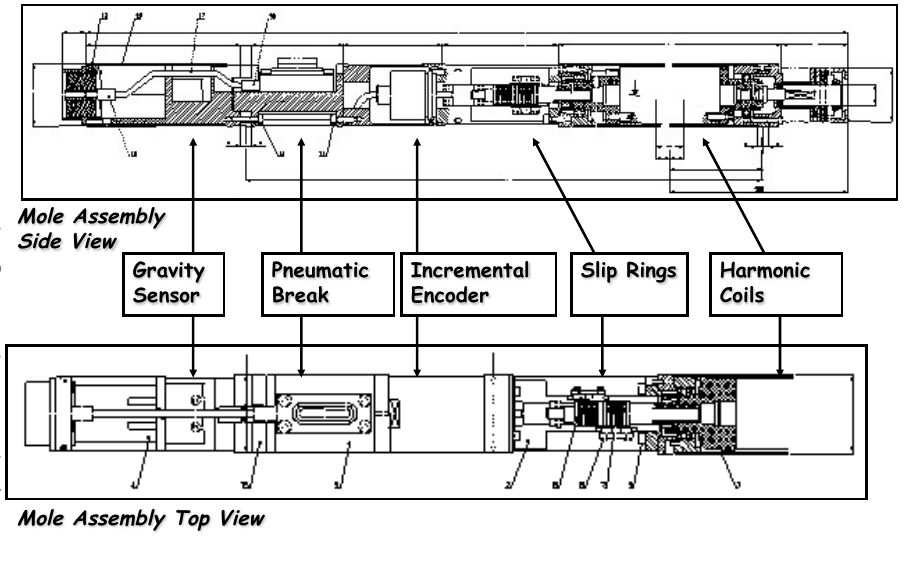}  
\includegraphics[height=4.2cm,trim=0 0.1cm 0 0]{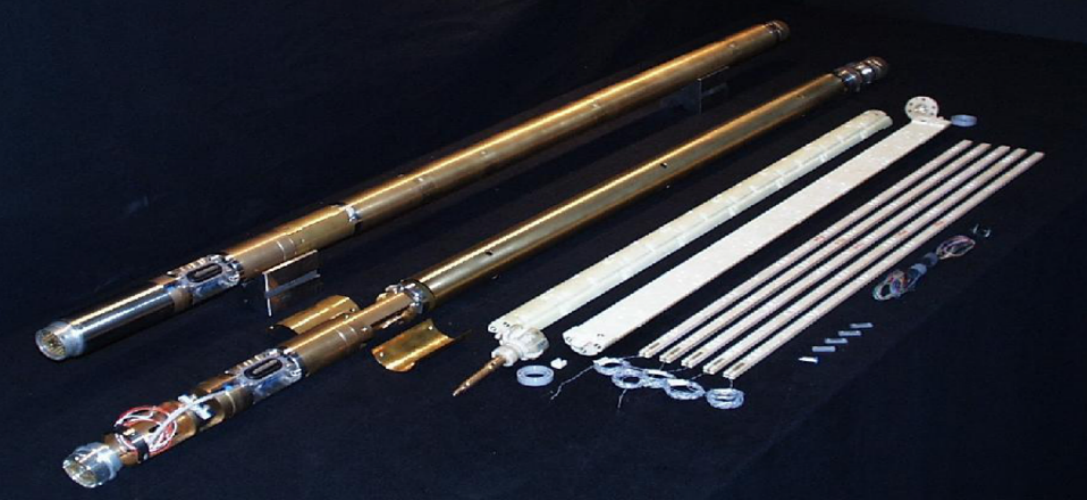}  
\includegraphics[height=4.2cm,trim=0 0.1cm 0 0]{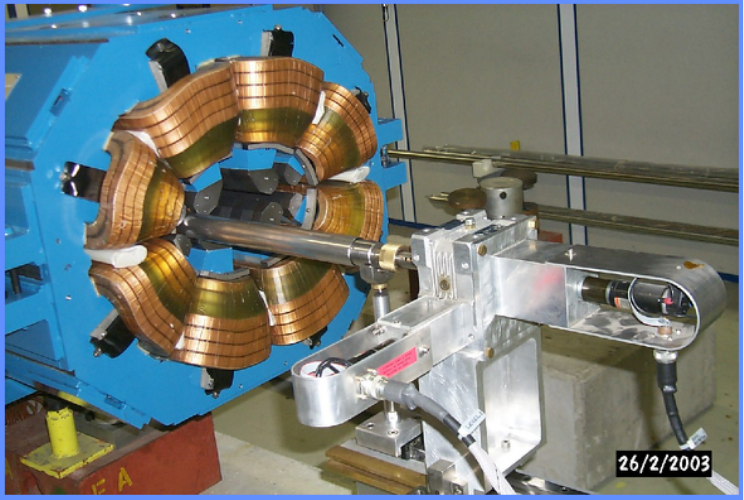}  
\caption{Top-left: Schematic drawing of a radial rotating coil in a magnetic field; top-right: Drawing of a rotating coil mole; bottom-left: photo of the components of a rotating coil mole; bottom-right: A rotating coil mole measurement in the MQW magnet. }
\label{fig:magmeas}
\end{center}
\end{figure}

To measure the magnetic field with sufficient precision to be relevant for the required field quality for circular accelerators, absolute field and field harmonics have to be measured with a precision of at least $10^{-4}$. Several magnetic measurements techniques can be applied to achieve this, e.g.:
\begin{itemize}
\item Rotating coils. \\
Used to measure multipoles and integrated field or gradient in all magnets.
\item Stretched wire. \\ 
To determine the magnetic centre and integrated gradient for n > 1 magnets.
\item Hall probes. \\
To make a field map.
\item Fixed pickup coils. \\
Measure the field on a current ramp.
\end{itemize}
Magnetic measurements are highly specialised technology and separate CAS schools are offered on this subject.
In Fig.~\ref{fig:magmeas} we can find some illustrations of measurements with the rotating coil technique. 

\section{Permanent and hybrid magnets}
\begin{figure}[ph!]
\begin{center}
\includegraphics[height=3.8cm]{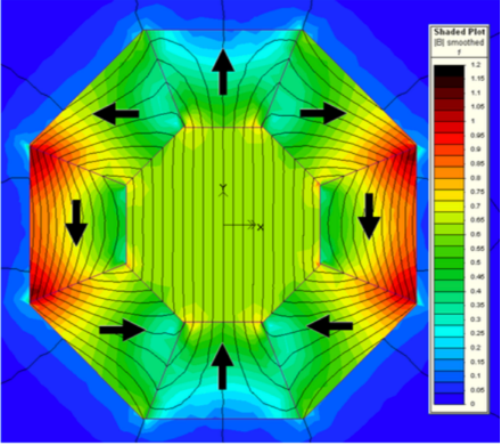}  \hspace{0.5cm}
\includegraphics[height=3.8cm,trim=0 0.7cm 0 0.3cm]{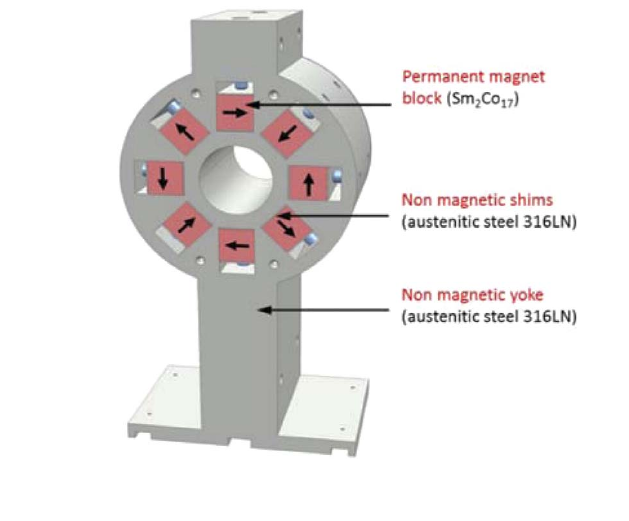}  \\
\includegraphics[height=3.5cm,trim=0 0.2cm 0 0]{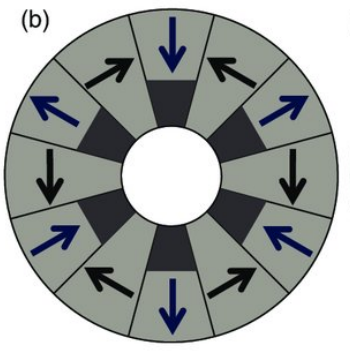}  \hspace{0.5cm}
\includegraphics[height=3.5cm]{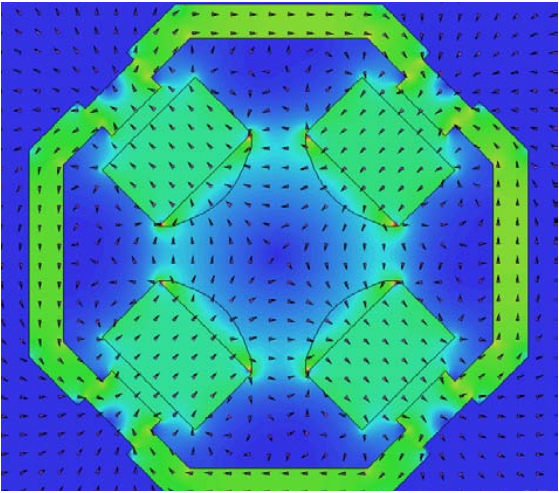} 
\caption{Top-left: flux pattern for a permanent magnet dipole; top-right: Permanent magnet dipole for Linac4 at CERN; bottom-left: Halbach array for a sextupole; bottom-right: Permanent magnet quadrupole with a steel yoke.}
\label{fig:permmag}
\end{center}
\end{figure}

Permanent magnetic materials can be used in various layouts to produce accelerator magnets. In Fig.~\ref{fig:permmag} a few examples of layouts with permanent magnets can be found. Permanent magnet blocks can be arranged to provide any multipolar field.

\begin{figure}[h!]
\begin{center}
\includegraphics[height=4.3cm,trim=0.5cm 0.5cm 0.5cm 0.0cm]{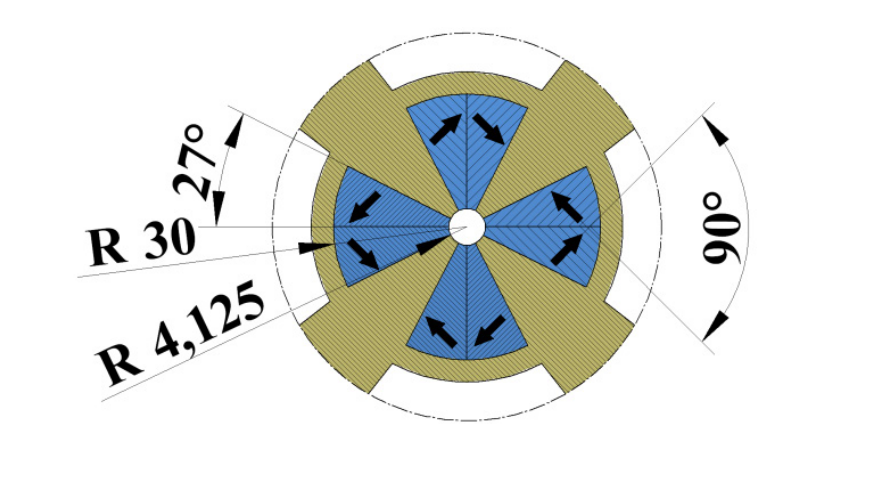}  
\includegraphics[height=4.3cm]{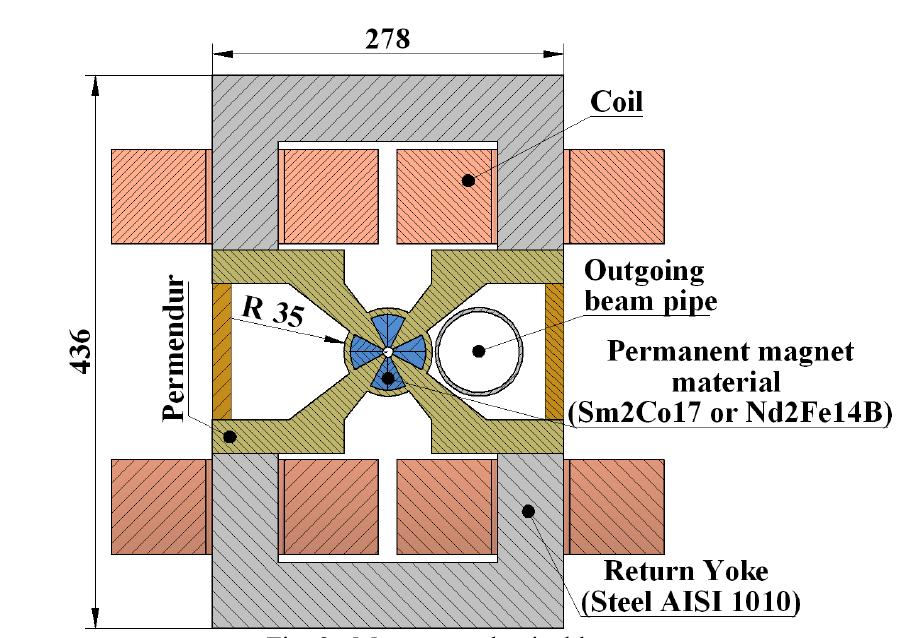}  \\
\includegraphics[height=4.5cm,trim=0 0.1cm 0 0]{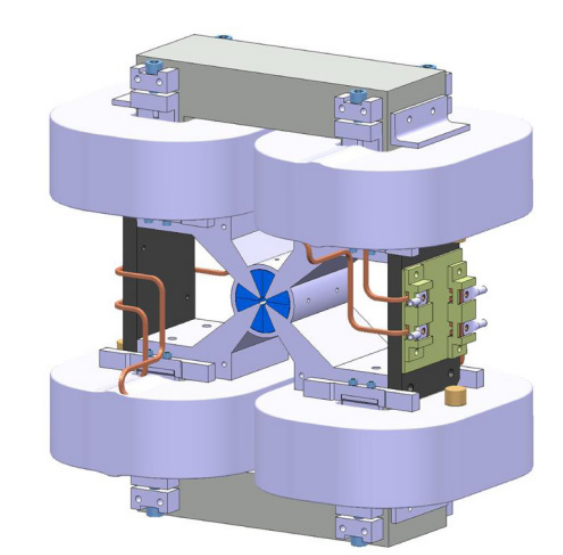}  
\caption{Hybrid quadrupole for the CLIC linear collider interaction area; top-left: permanent magnet part; top-right: cross section; bottom: prototype magnet.}
\label{fig:hybrid}
\end{center}
\end{figure}

In order to make tunable magnets with permanent magnet materials, it is possible to make hybrid solutions. In Fig.~\ref{fig:hybrid} we can see a quadrupole designed for the interaction region of the CLIC linear collider (CERN) with a gradient of  $530 \UT/\UmZ$ in a aperture of $8.25 \Umm$ diameter that is tuneable in the range $10\%-100\%$.

\section{Examples of magnets over the history of accelerators}
\begin{figure*}[h!]
\begin{center}
\includegraphics[height=6cm]{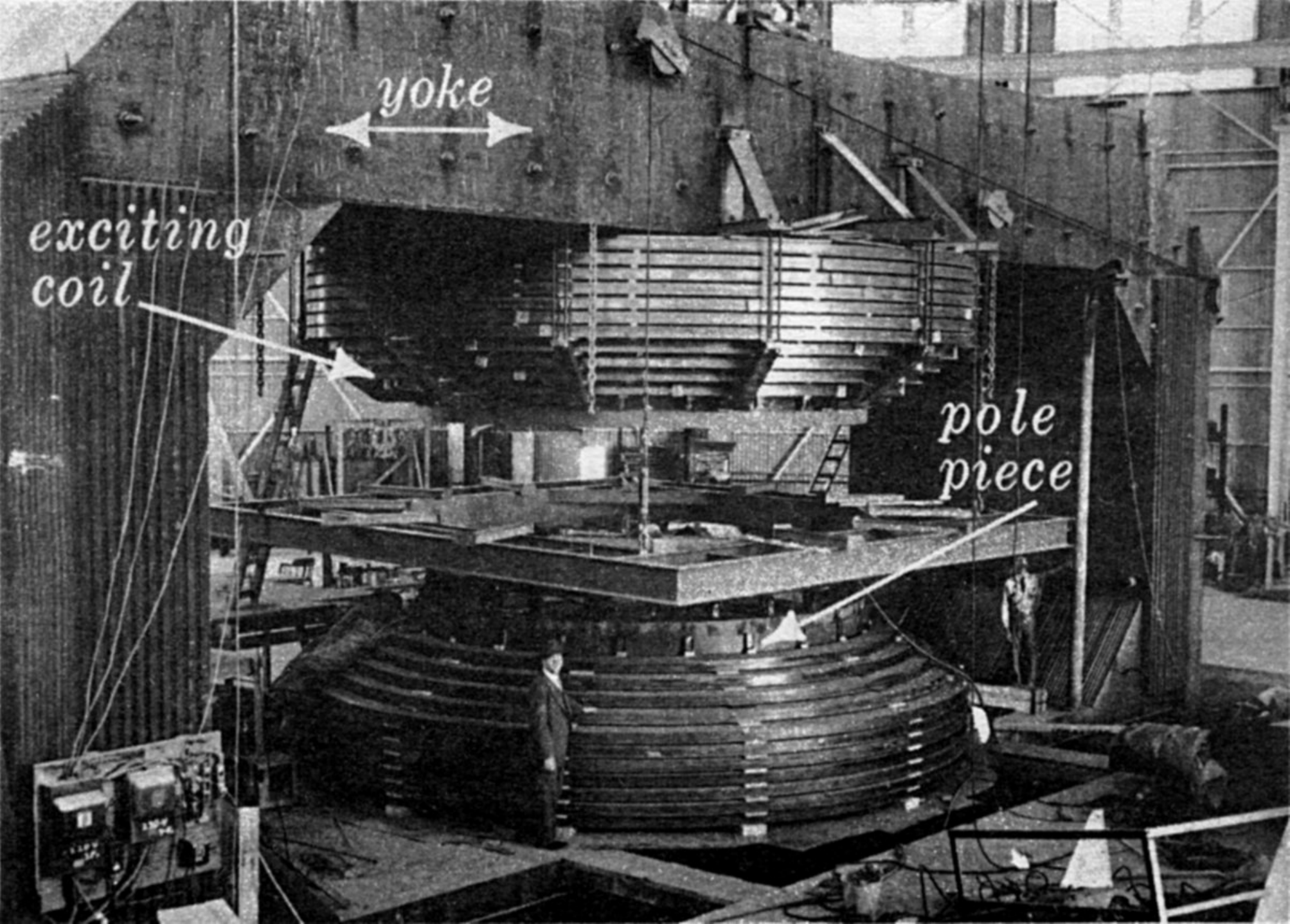} 
\includegraphics[height=6cm]{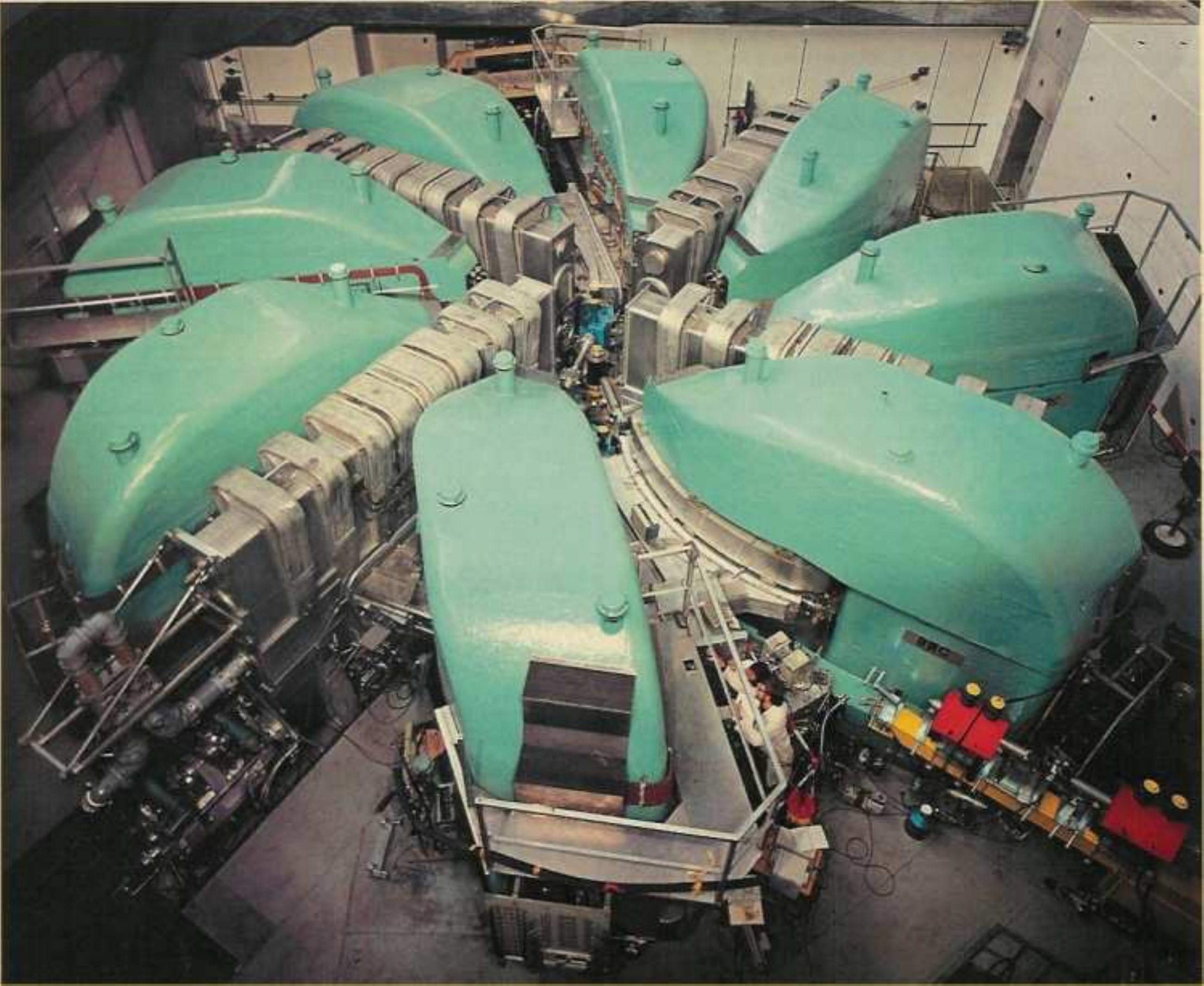} 
\caption{Left: The Berkeley 184" cyclotron at Berkeley; right: The PSI sector cyclotron of 1974.}
\label{fig:cyclotrons}
\end{center}
\end{figure*}

Practical accelerators started in 1929-1930 with the cyclotron, invented by Ernest O. Lawrence at the University of California, Berkeley. Cyclotrons developed very fast and machines with single and separated magnets have been built, like the sector cyclotrons; see Fig.~\ref{fig:cyclotrons}.

\begin{figure}[h!]
\begin{center}
\includegraphics[height=5.0cm,trim=0 0.2cm 0 0]{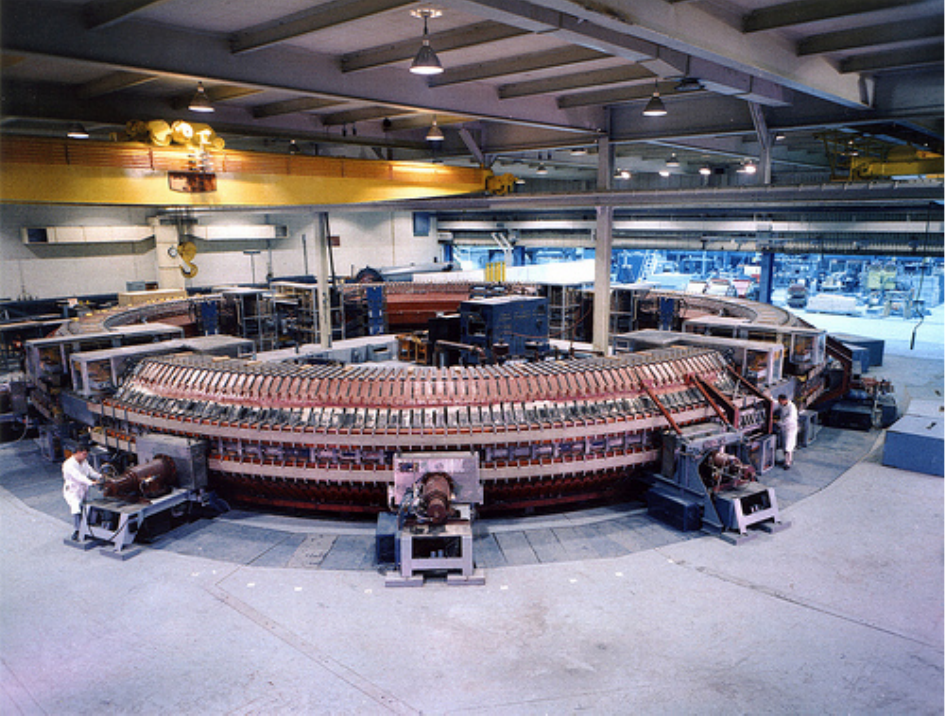}  
\includegraphics[height=5.0cm,trim=0 0.2cm 0 0]{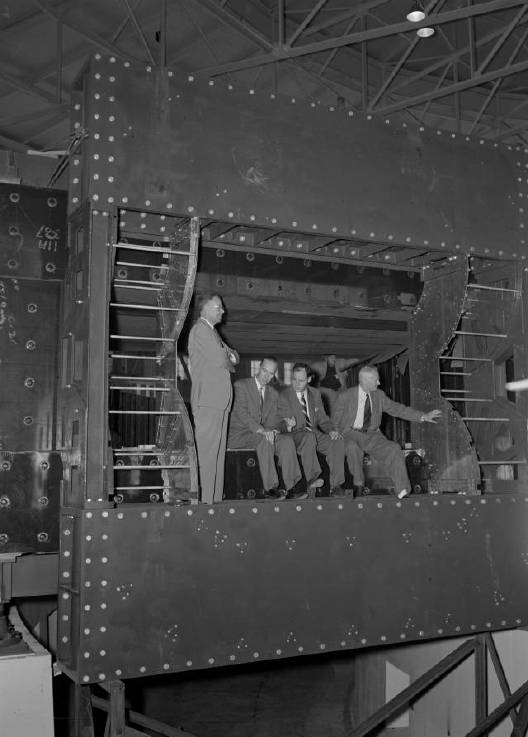}
\includegraphics[height=5.0cm,trim=0 0.2cm 0 0]{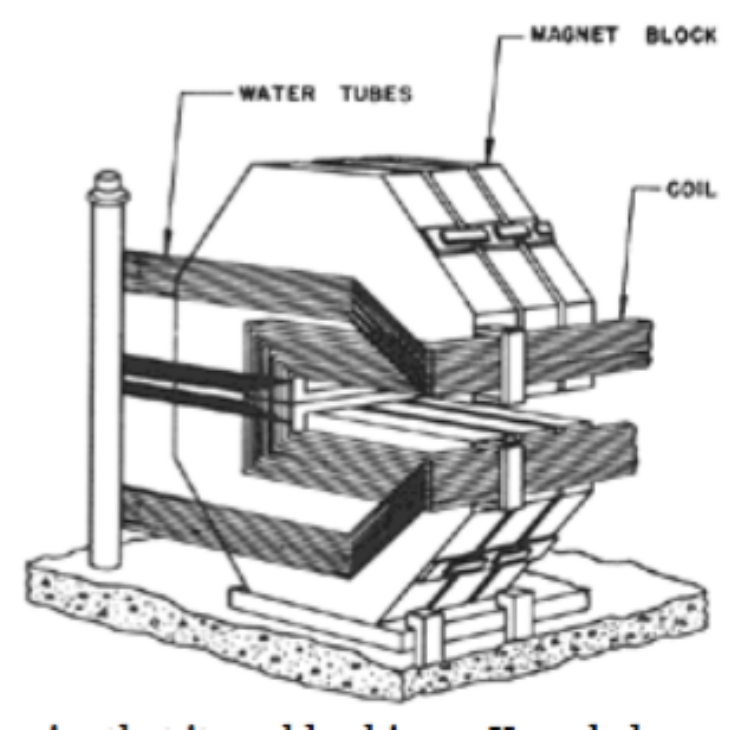}
\caption{Left: The cosmotron at BNL (1953); middle: people sitting in the huge aperture of a magnet for a weak focusing synchrotron; right: schematic drawing of a magnet for a weak focusing synchrotron.}
\label{fig:weakfocus}
\end{center}
\end{figure}

Synchrotrons were invented in the beginning of the 1950-s featuring a string of magnets in a ring shaped accelerator. The Cosmotron at Brookhaven (US) was commissioned in 1953 and had a beam energy of $3.3 \UGeV$. It was a weak focussing machine and thus had a very large aperture of 20$\Ucm$ x 60$\Ucm$ (Fig.~\ref{fig:weakfocus}).

\begin{figure}[hb!]
\begin{center}
\includegraphics[height=5cm]{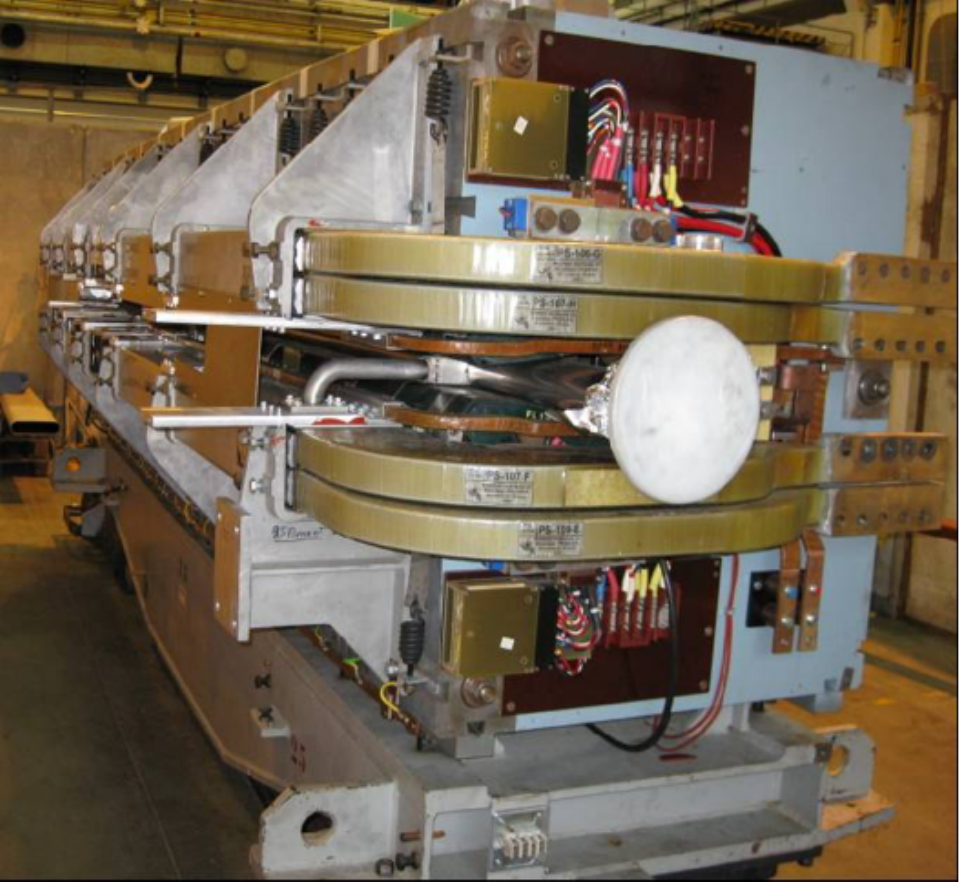} 
\includegraphics[height=5cm]{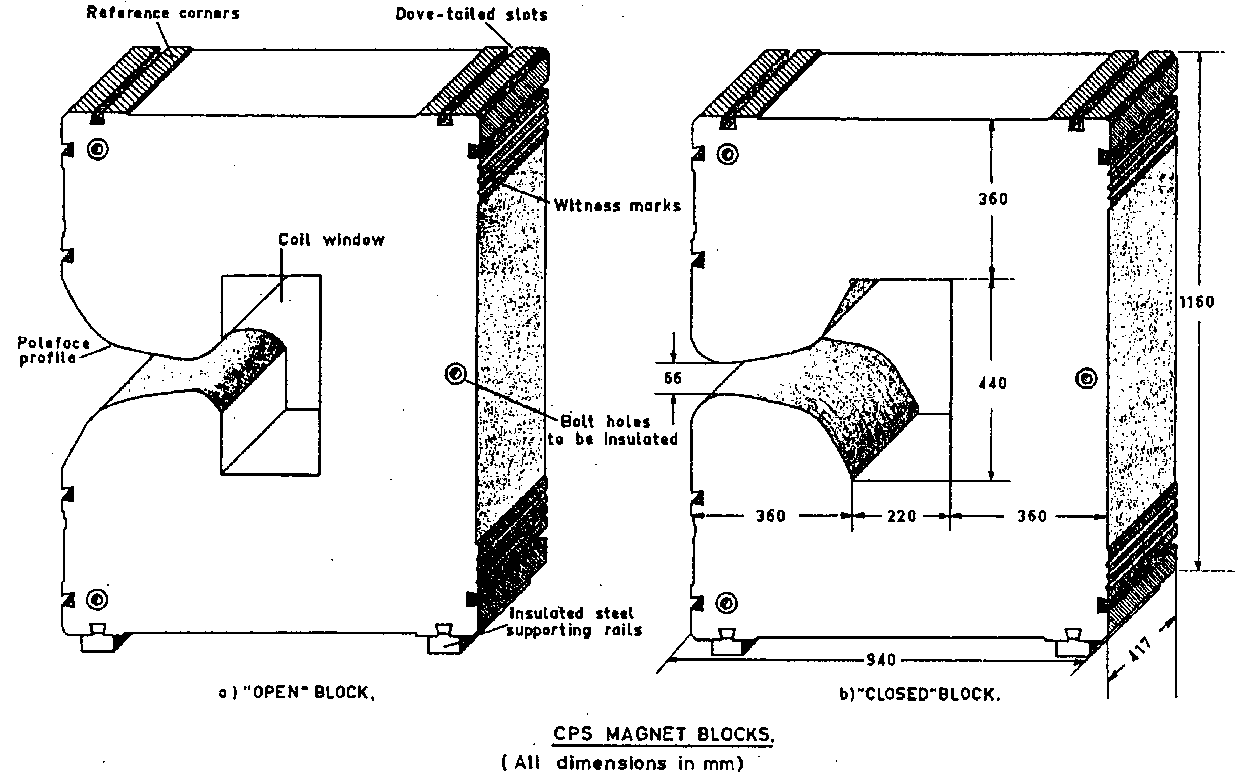}\\
\includegraphics[height=5.5cm,trim=0 0.4cm 0 0.cm]{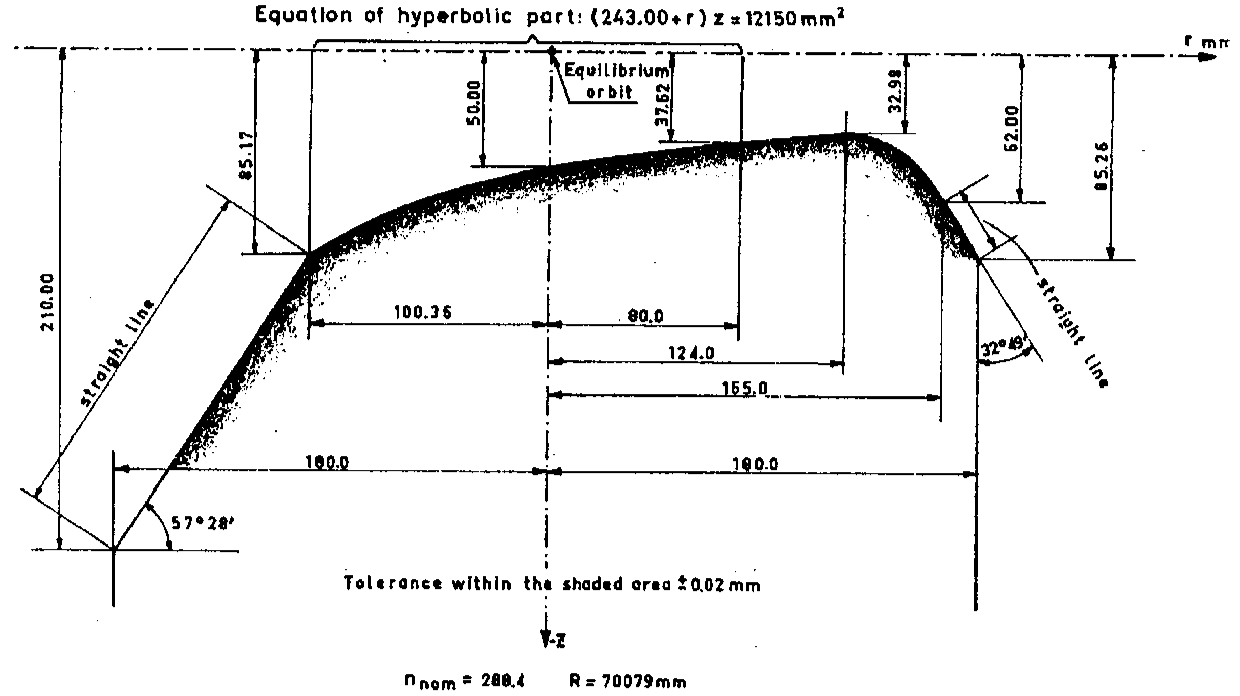}
\caption{Top-left: Combined function magnet for the strong focusing Proton Synchrotron (PS) at CERN; top-right: two types of focusing magnet blocks; bottom: PS magnet pole.}
\label{fig:PS}
\end{center}
\end{figure}

\vspace*{-5mm}Beam size and thus aperture size could be reduced by close to an order of magnitude by employing strong focusing. The first strong focusing synchrotrons used combined function magnets. An example is the Proton Synchrotron at CERN that was commissioned in 1959. It reached $24.3 \UGeV$ at an average flux density of $1.2\UT$ in a combined function magnet (Fig.~\ref{fig:PS}).

\begin{figure}[h!]
\begin{center}
\includegraphics[height=5cm]{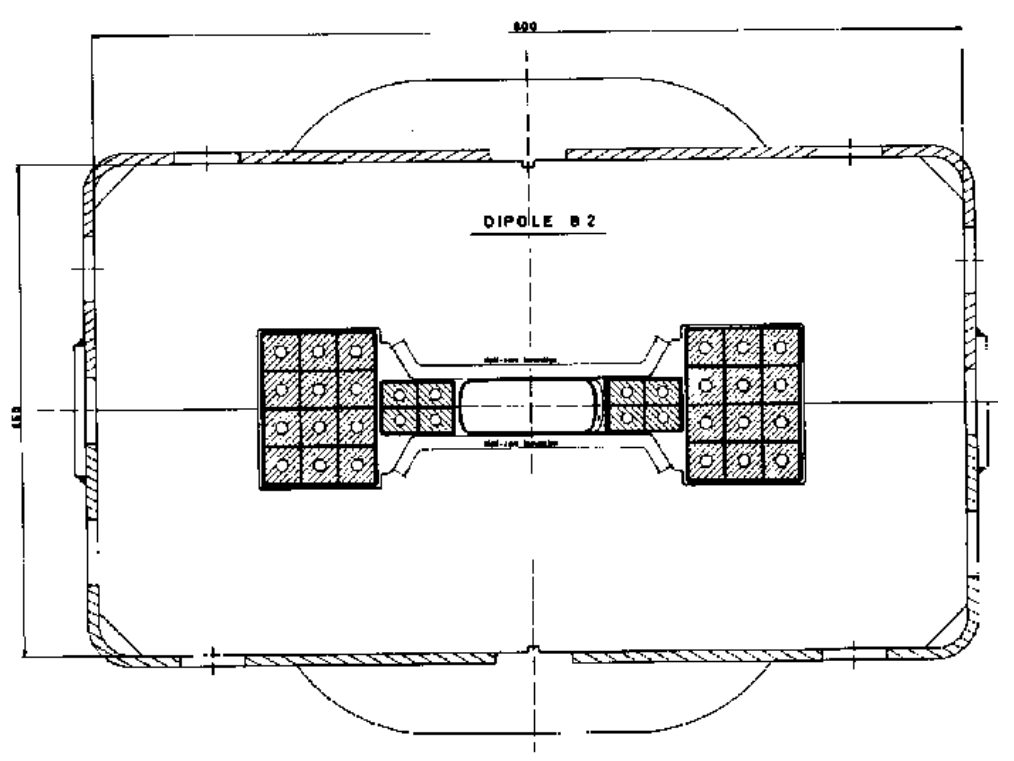} 
\includegraphics[height=5cm]{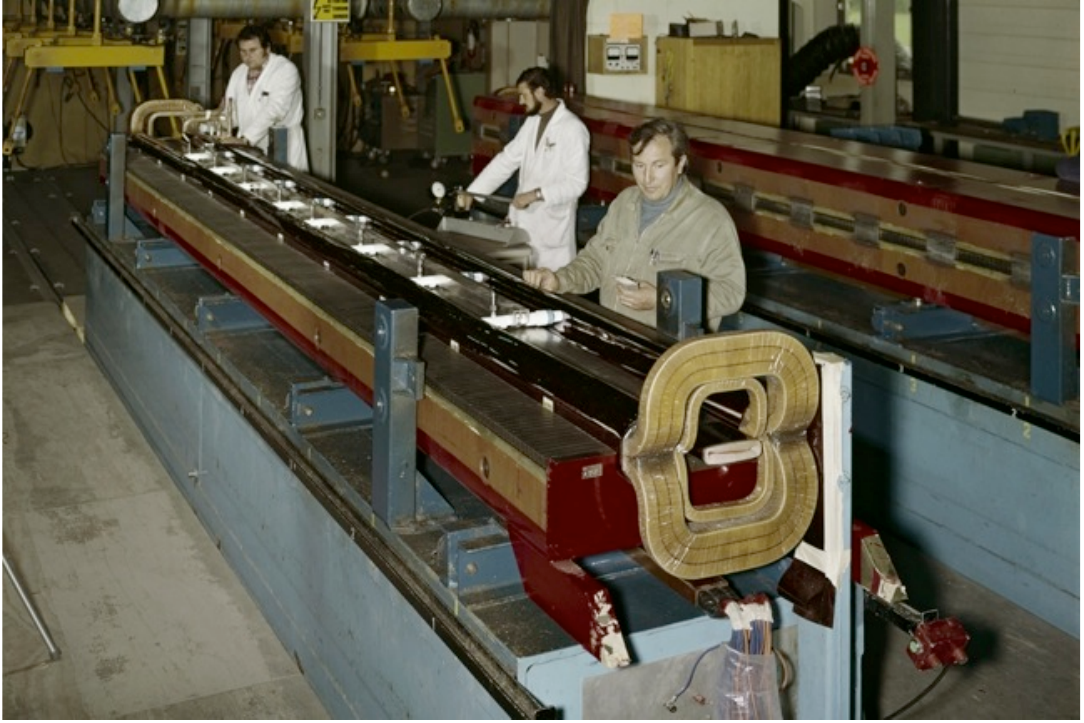}  \\
\includegraphics[height=5cm]{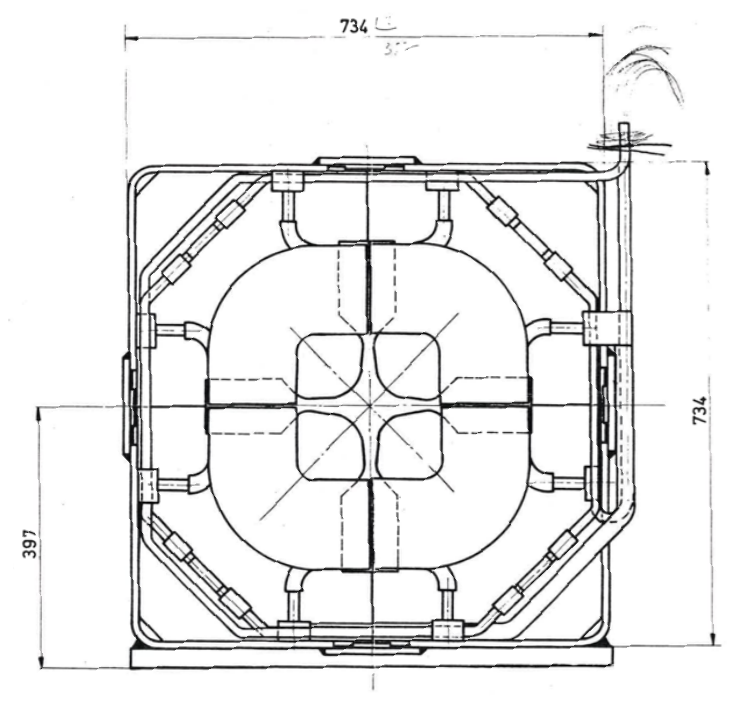} 
\includegraphics[height=5cm]{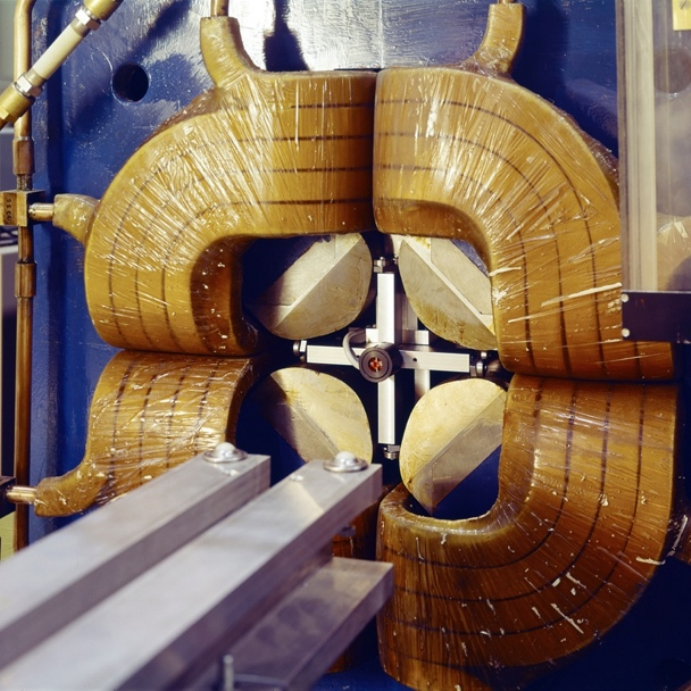}
\caption{The SPS main magnets: top-left: dipole cross-section; top-right: a photograph taken during assembly of the dipoles in the early 1970s; bottom-left: quadrupole cross-section; bottom-right: detail of the SPS quadrupole.}
\label{fig:spsdip}
\end{center}
\end{figure}

The next step in the development of synchrotrons is the strong focusing synchrotrons with separated function magnets. The Super Proton Synchrotron at CERN is a nice example of this technology. In Fig.~\ref{fig:spsdip} one can see the SPS magnets. The dipole magnet generates a flux density $B_{\rm max} = 2.05 \UT$ using  a 16-turn coil with $I_{\rm max}= 4900\UA$ in a $52\Umm$ high and $92\Umm$ wide aperture and the quadrupole provides a gradient of $20.7 \UT/\UmZ$ with a coil of 16 turns and a current of $1938 \UA$ in an aperture with a $44\Umm$ radius. The machine accelerates protons up to an energy of $450\UGeV$.

\begin{figure}[h!]
\begin{center}
\includegraphics[height=5cm]{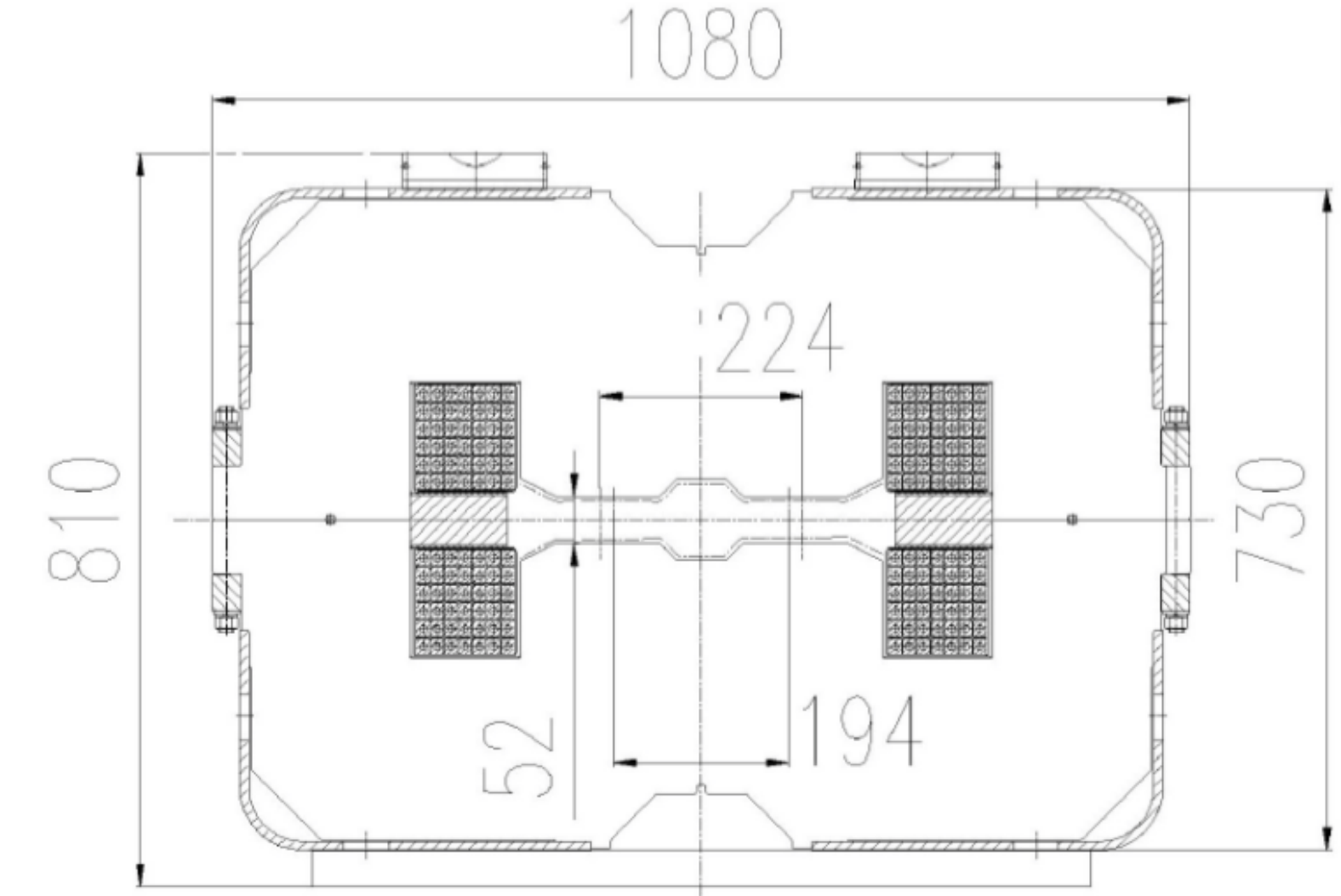} 
\includegraphics[height=5cm]{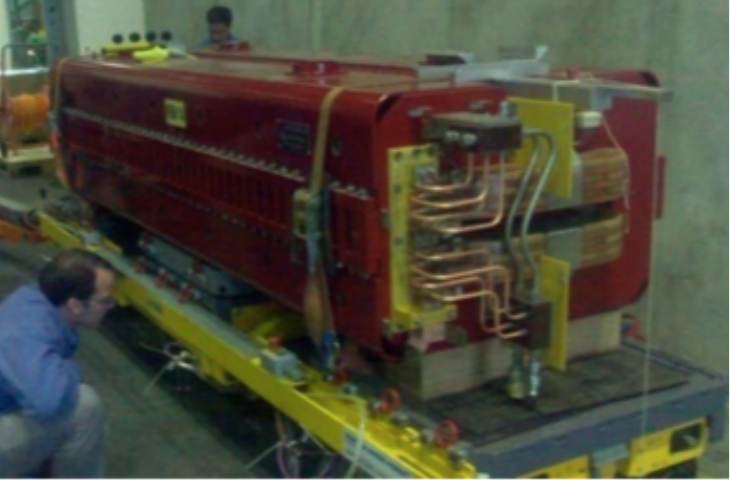}
\caption{Recent warm magnet for the LHC: left: MBW separation dipole cross section; right: MBW during installation.}
\label{fig:lhcmag}
\end{center}
\end{figure}

\vspace*{-3mm}A more recent example of a warm magnet can be found in Fig.~\ref{fig:lhcmag}. The MBW separation dipole is used in the LHC to further separate the beams by $30 \Umm$ in the collimation areas; it has a field of $1.42\UT$ in a $52\Umm$ pole gap. 

\begin{figure*}[h!]
\begin{center}
\includegraphics[height=5cm]{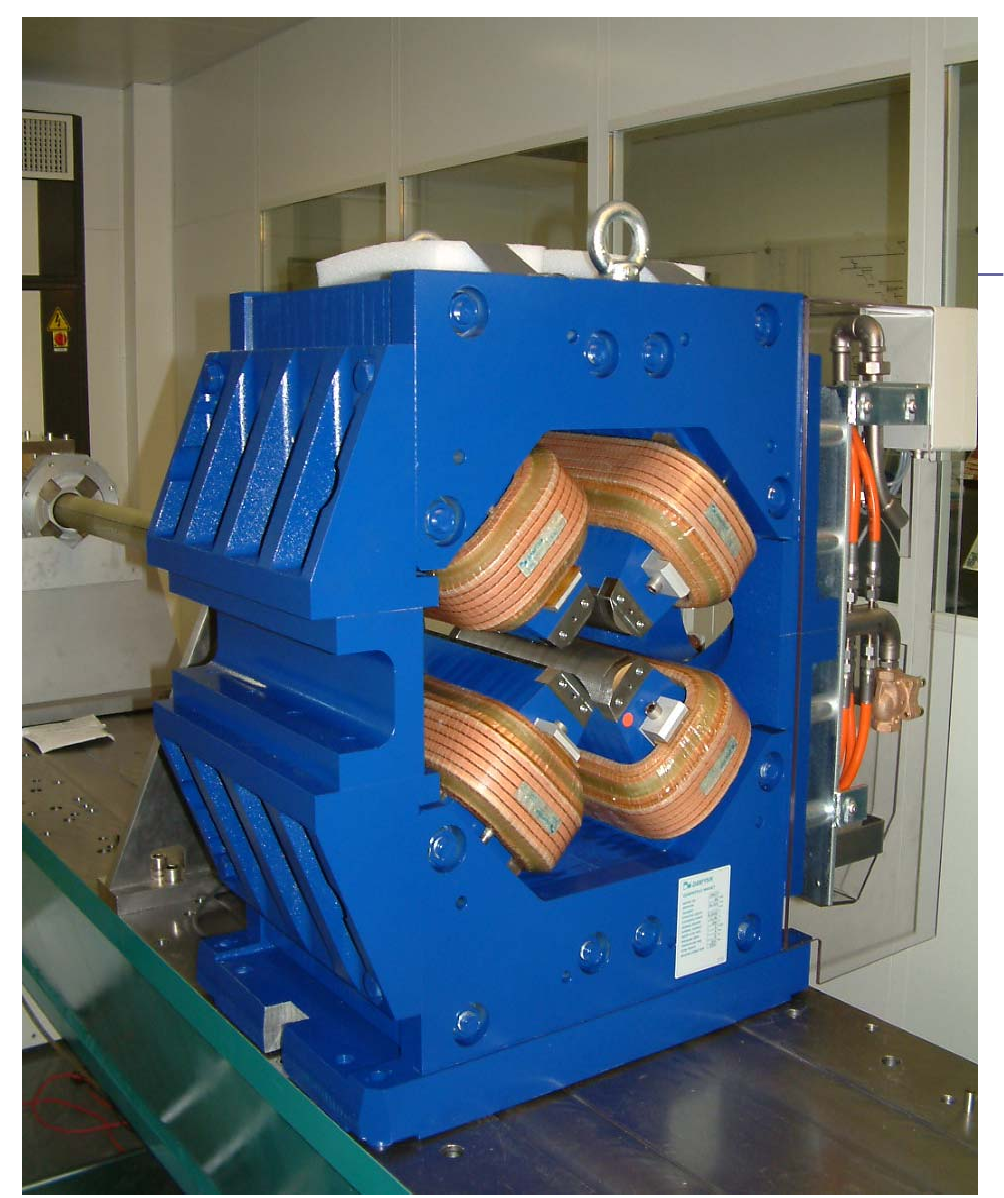} \hspace{ 0.1cm}
\includegraphics[height=5cm]{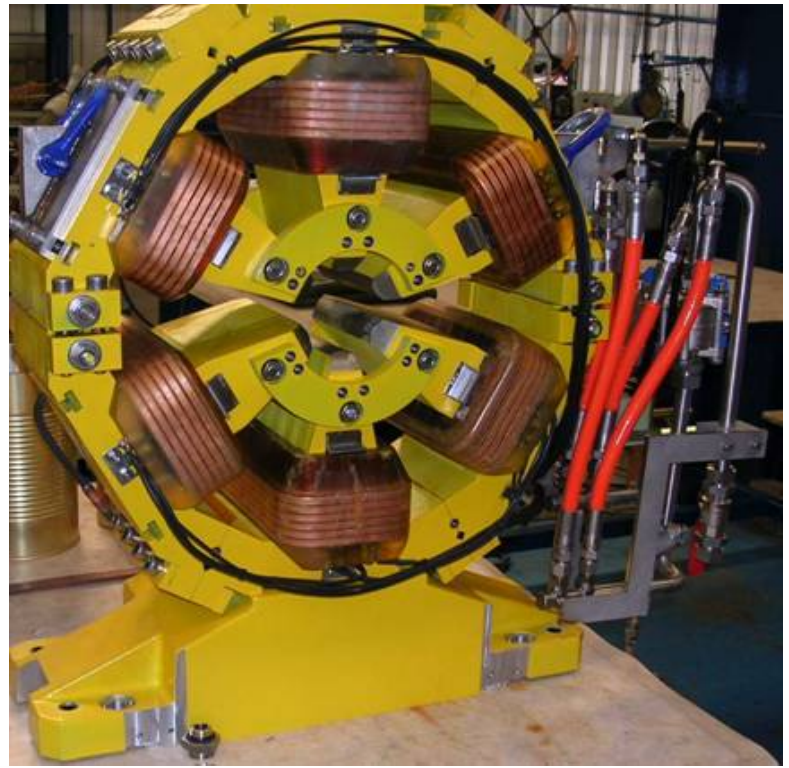}   \hspace{ 0.1cm}
\includegraphics[height=5cm]{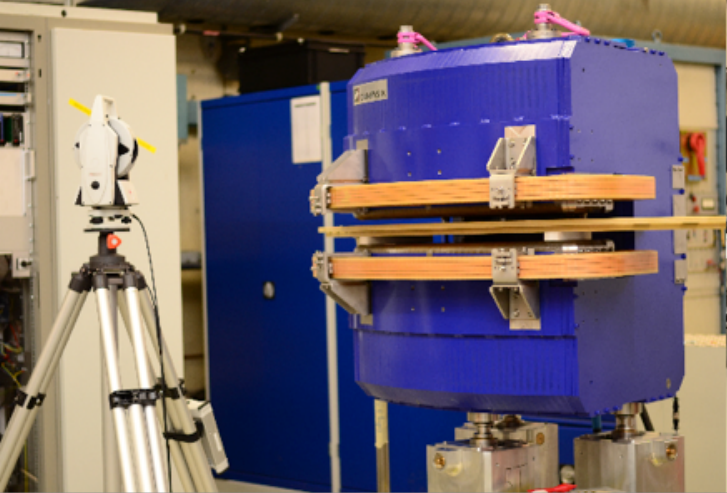} \hspace{ 0.1cm}
\includegraphics[height=5cm]{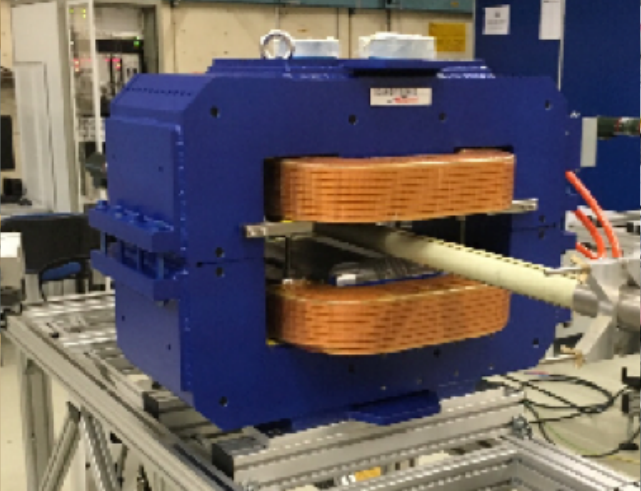}
\caption{Top-left: SOLEIL quadrupole; top-right: SOLEIL sextupole; bottom-left: ELENA ring dipole; bottom-right: ELENA TL dipole.}
\label{fig:smallrings}
\end{center}
\end{figure*}

New elegant designs have recently been done for synchrotron light sources and small storage rings. In Fig.~\ref{fig:smallrings} we can find a dipole and a sextupole of the SOLEIL synchrotron light source that is located near Gif-sur-Yvette south of Paris. We can see as well two dipoles for the ELENA antiproton decelerator ring at CERN.

\section*{Acknowledgements}
For this lecture I used material from lectures, seminars and reports from many colleagues. Special thanks goes to:
Davide Tommasini, Attilio Milanese, Antoine Dael, Stephan Russenschuck and Thomas Zickler, and to the people who taught me, years ago, all the fine details about magnets.
 
\section*{Bibliography}
Books: \\
G.E.Fisher, \emph{Iron Dominated Magnets}, AIP Conf. Proc., 1987 -- Volume 153, pp. 1120-1227 \\
J. Tanabe, \emph{Iron Dominated Electromagnets}, World Scientific, ISBN 978-981-256-381-1, May 2005 \\
P. Campbell, \emph{Permanent Magnet Materials and their Application}, ISBN-13: 978-0521566889 \\
S. Russenschuck, \emph{Field computation for accelerator magnets: analytical and numerical methods for electromagnetic design and optimization}, Wiley, 2010, ISBN-13: 978-3527407699 \\
Schools: \\
D. Brandt, et al., \emph{CAS Magnets, Bruges, 2009}, CERN-2010-004 \\
D. Einfeld, \emph{CAS Introduction to Accelerator Physics, Frascati 2008, Magnets (Warm)}, \\ http://cas.web.cern.ch/schools/frascati-2008 \\
D. Tommasini, \emph{CAS Introduction to Accelerator Physics, Varna 2010, Magnets (Warm)}, \\ http://cas.web.cern.ch/schools/varna-2010 \\
R. Bailey, at al., \emph{CAS Power Convertors, Baden 2014}, CERN-2015-003 \\
S. Turner, et al., \emph{CAS Measurement and Alignment of Accelerator and Detector Magnets, Anacapri 1997}, CERN-98-05 \\
Reports:\\
D. Tommasini,  \emph{Practical definitions and formulae for magnets}, CERN,Tech. Rep. EDMS 1162401, 2011 \\
\end{document}